\documentclass[12pt]{iopart}
\usepackage[T1]{fontenc}
\usepackage[latin9]{inputenc}
\usepackage{geometry}
\usepackage{fancyhdr}
\usepackage{color}
\usepackage{graphicx}
\usepackage{float}
\usepackage{setspace}
\usepackage[font={small}]{caption}

\usepackage{iopams}  
\usepackage{subfig}
\makeatother

\begin{document}

\title[Coarsening dynamics in ZRP]{Coarsening dynamics in condensing zero-range processes and size-biased birth death chains}

\author{Watthanan Jatuviriyapornchai, Stefan Grosskinsky}

\address{Mathematics Institute, University of Warwick, Coventry CV4 7AL, UK}
\ead{w.jatuviriyapornchai@warwick.ac.uk, s.w.grosskinsky@warwick.ac.uk}
\vspace{10pt}
\begin{indented}
\item[]January 2016
\end{indented}

\begin{abstract}
Zero-range processes with decreasing jump rates are well known to exhibit a condensation transition under certain conditions on the jump rates, and the dynamics of this transition continues to be a subject of current research interest. Starting from homogeneous initial conditions, the time evolution of the condensed phase exhibits an interesting coarsening phenomenon of mass transport between cluster sites characterized by a power law. We revisit the approach in [C.\ Godr\`eche, J.\ Phys.\ A: Math.\ Gen., \textbf{36}(23) 6313 (2003)] to derive effective single site dynamics which form a non-linear birth death chain describing the coarsening behaviour. We extend these results to a larger class of parameter values, and introduce a size-biased version of the single site process, which provides an effective tool to analyze the dynamics of the condensed phase without finite size effects and is the main novelty of this paper. Our results are based on a few heuristic assumptions and exact computations, and are corroborated by detailed simulation data.
\end{abstract}

\section{Introduction}
Under certain conditions on the jump rates, stochastic particle systems
can exhibit a condensation transition where a non-zero fraction of
all particles accumulates in a condensate, provided the particle density
exceeds a critical value. This has been studied for various models
with homogeneous factorized stationary distributions (see e.g. \cite{chleboun2014condensation,evans2014condensation}
for recent summaries), including zero-range processes of the type
introduced in \cite{drouffe1998simple,evans2000phase}, inclusion
processes with a rescaled system parameter \cite{grosskinsky2013dynamics,cao2014dynamics}
and explosive condensation models \cite{waclaw2012explosive,chau2015explosive}. The role
of spatial inhomogeneities and their interplay with particle interactions
is summarized in detail for zero-range processes in \cite{godreche2012condensation}
(see also \cite{chleboun2014condensation} and \cite{mailler2015condensation} for further references),
and in this paper we focus on spatially homogeneous processes. 

\begin{figure}[t]
\centering
\subfloat[Coarsening]{\protect\includegraphics[scale=0.65]{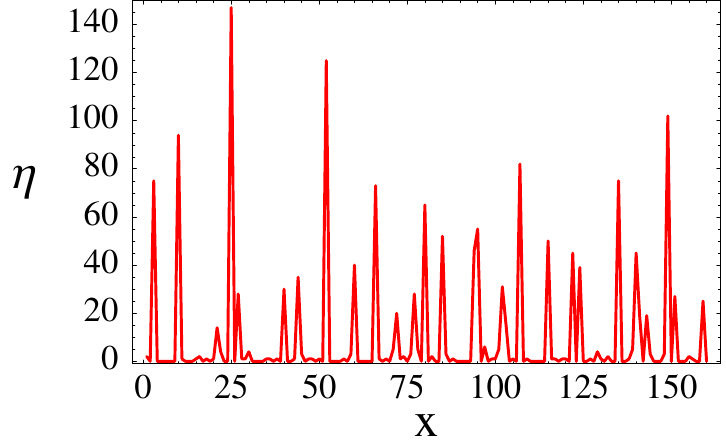}}\subfloat[Saturation]{\protect\includegraphics[scale=0.65]{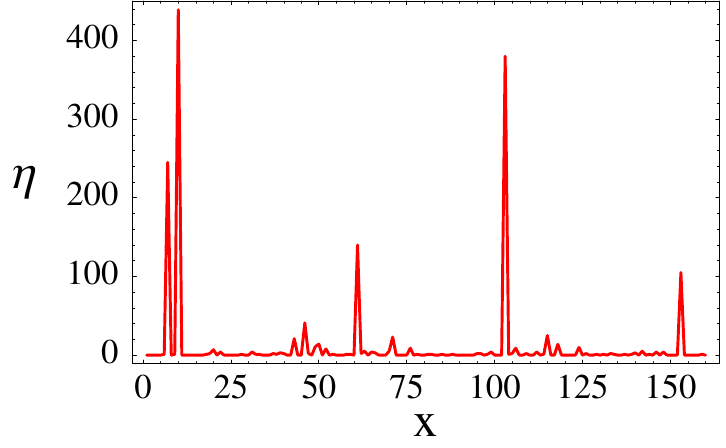}}\subfloat[Stationarity]{\protect\includegraphics[scale=0.65]{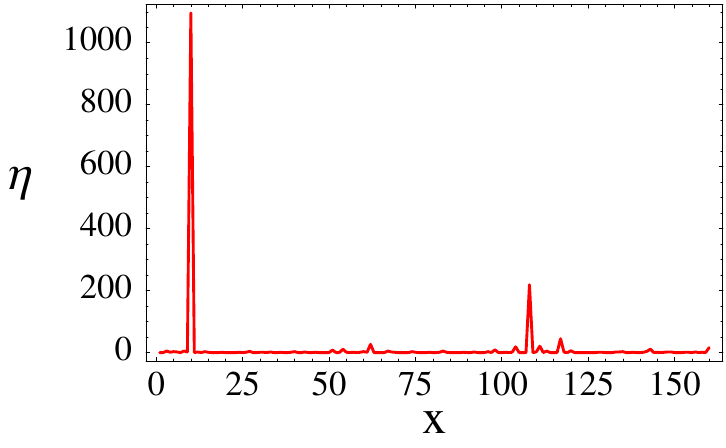}
}
\protect\caption{\label{fig:fig1dynamic}
Illustration of the condensation dynamics via sample density profiles for the symmetric zero-range process with homogeneous initial conditions. In the coarsening regime (a) clusters exchange particles and large ones grow at the expense of small ones. Clusters do not change location, and with only few cluster sites left the system saturates (b). This leads to the stationary state with a single condensate (c) which changes position only on a slower time scale. Note the different ranges of the $y$-axes.}
\end{figure}

While the stationary behaviour of condensing systems with factorized
distributions has been understood in great detail also on a rigorous
level \cite{chleboun2014condensation,evans2000phase,jeon2000size,godreche2003dynamics,grosskinsky2003condensation,armendariz2009thermodynamic,armendariz2013zero},
the dynamics of these processes continue to attract research interest.
The zero-range processes introduced in \cite{drouffe1998simple,evans2000phase}
with jump rates of the form $g(k)=1+b/k^{\gamma},\;b>0,\;\gamma\in(0,1]$
for occupation numbers $k\geq1$, exhibit particularly rich dynamic
phenomena involving several time scales which have been summarized
in \cite{godreche2005dynamics} for the case $\gamma=1$. The fastest
is the usual hydrodynamic time scale of order $L$ for asymmetric
and order $L^{2}$ for symmetric systems of size $L$ on which non-stationary
density profiles evolve in time, which has been studied heuristically
in \cite{schutz2007hydrodynamics} with partial rigorous results in
\cite{stamatakis2015hydrodynamic}. Within that scale, the excess mass
in supercritical systems has accumulated in isolated cluster sites which
form the condensed phase. These cannot be included in a hydrodynamic
description via density fields, but rather act as boundary reservoirs
scattered at random positions in the system. The sites exchange particles
on a slower time scale identified as $L^{2}$ (asymmetric) and $L^{3}$ (1D symmetric) in \cite{godreche2003dynamics,godreche2005dynamics}
for $\gamma=1$, with related recent rigorous progress in \cite{beltran2015martingale}.
These dynamics in the condensed phase correspond to an interesting
coarsening process with a decreasing number of increasingly large
cluster sites illustrated in Figure \ref{fig:fig1dynamic}. Typical observables follow
a power law time evolution, corresponding to a self-similar process
on an infinite lattice, which exists only in a scaling window on finite
systems before reaching stationarity with a single cluster
or condensate remaining. The stationary dynamics of this condensate form the longest time scale,
have first been identified in \cite{godreche2005dynamics} on the
scale $L^{b}$ (asymmetric and complete graph) and $L^{1+b}$(1D symmetric), and are now
understood on a rigorous level \cite{beltran2010tunneling,beltran2012metastability,landim2014metastability,armendariz2015metastability}. Related recent results on the equilibration dynamics of condensation phenomena in population genetics can be found in \cite{dereich2013emergence}.

In this paper, we revisit the analysis of the coarsening dynamics in
\cite{godreche2003dynamics,godreche2005dynamics}. We provide a generic
derivation for the single-site dynamics as non-linear birth death
chains, and study the coarsening scaling law for values of the parameter
$\gamma$ extending to the interval $(0,1]$. The main novelty is a direct description of the size-biased version of the single-site dynamics, which provides an interesting tool to analyze the coarsening dynamics without any finite size effects. It is also numerically more efficient than previous approaches since the volume of the condensed phase does not vanish in time as in the zero-range process or the direct mapping to birth death chains. 
We focus the presentation of our results on processes on complete graphs, and discuss the role of other homogeneous geometries in the conclusion. In the next section, we introduce the model and summarize background results. Our description of the coarsening dynamics and related results are given in Sections 3 and 4, and we conclude in Section 5.

\section{Zero-range processes and condensation}

\subsection{Definition and notation}

The zero-range process $(\boldsymbol{\eta}(t):t>0)$ is a stochastic
particle system defined on a lattice $\Lambda$ which is a discrete set of sites that can be finite of size $L$ or infinite. Configurations
are denoted by $\boldsymbol{\eta}=(\eta_{x}:x\in\Lambda)\in\Omega$,
where $\eta_{x}\in\mathbb{N}$ is the number of particles at site
$x\in\Lambda$ with the state pace $\Omega=\mathbb{N}^{\Lambda}$. Let $q(x,y)\geq0$ be the transition probabilities
of an irreducible random walk on $\Lambda$ with $q(x,x)=0$. A particle
at site $x$ jumps to site $y$ with rate $q(x,y)g(\eta_{x})$, which
depends only on the occupation number on the departure site $x$ through
a non-negative function $g$, for which we assume that $g(n)=0$ if
and only if $n=0$. 
The dynamics of the process is then conveniently characterized by the infinitesimal generator
\begin{equation}
(\mathcal{L}f)(\boldsymbol{\eta})=\sum_{x,y\in\Lambda}g(\boldsymbol{\eta}_{x})q(x,y)(f(\boldsymbol{\eta}^{x\rightarrow y})-f(\boldsymbol{\eta})),\label{eq:GenZRP}
\end{equation}
describing the expected change of suitable test functions $f:\;\Omega\to\mathbb{R}$
(see e.g. \cite{andjel1982} for details). Here we have used the standard
notation $\boldsymbol{\eta}^{x\rightarrow y}$ for the configuration where one
particle has jumped from site $x$ to site $y$, i.e. $\eta_{z}^{x\rightarrow y}=\eta_{z}-\delta_{z,x}+\delta_{z,y}$, where $\delta$ denotes the Kronecker delta. The adjoint operator $\mathcal{L}^*$
with respect to the distribution $\pi_t$ at time $t$ characterizes the right-hand
side of the usual master equation $\frac{d}{dt} \pi_t (\eta )=(\mathcal{L}^* \pi_t )(\eta )$. 
We do not give this in detail and use the generator (\ref{eq:GenZRP}) instead, since it is more convenient for our purposes.

For spatially homogeneous jump probabilities, the stationary distributions of the zero-range process are well known \cite{andjel1982,evans2005} to be of a product form with marginals 
\begin{equation}\label{eq:marg}
\nu_{\phi} [\eta_{x}=n]=\frac{1}{z(\phi)}w(n)\phi^{n},
\end{equation}
where $\phi\geq 0$ is the fugacity parameter controlling the particle density. The stationary weight is given by $w(n)=\prod_{k=1}^{n}g(k)^{-1}$ and the normalization by the single site partition function
\[
z(\phi):=\sum_{n=0}^{\infty}w(n)\phi^{n}\ .
\]
These distributions exist on finite or infinite lattices for all $\phi\geq 0$ such that $z(\phi )<\infty$, with the particle density
\[
R(\phi):=\sum_{n=0}^{\infty}n \nu_{\phi} [\eta_{x}=n]=\phi\partial_{\phi}\log(z(\phi)),
\]
a monotone increasing function of $\phi$ with $R(0)=0$.

\subsection{Condensation}

In this paper, we focus on zero-range processes introduced in \cite{drouffe1998simple,evans2000phase} with jump rates 
\begin{equation}\label{rates}
g(k) =  
\left\{ \begin{array}{cl}
0 & \mbox{if }k=0, \\
1+\frac{b}{k^\gamma} & \mbox{if }k\geq 1,
\end{array} \right.
\end{equation}
with parameters $b>0$ and $\gamma\in(0,1]$. It is known that this system exhibits condensation if $\gamma\in (0,1)$ or $\gamma =1$ and $b>2$. In this case, the stationary weights $w(n)$ have a subexponential decay as $n\to\infty$ and the distributions (\ref{eq:marg}) exist for all $\phi \in [0,1]$ with the maximal invariant measure $\nu_1$ and critical density $\rho_c =R(1)<\infty$. When the imposed particle density $\rho$ in a finite system of size $L$ exceeds $\rho_c$, the system phase separates into a homogeneous background distributed according to $\nu_1$ with critical
density, and the condensate where the excess mass $(\rho -\rho_c )L$ concentrates on a single randomly located lattice site. This phenomenon has been very well understood on the level of stationary distributions and the equivalence of ensembles in a series of papers \cite{chleboun2014condensation,evans2000phase,jeon2000size,godreche2003dynamics,grosskinsky2003condensation,armendariz2009thermodynamic,armendariz2013zero}.

The dynamics of this condensation phenomenon remain an area of active research interest with recent rigorous results in \cite{beltran2015martingale}, and also in the context of related population models \cite{dereich2013emergence}. The zero-range process with rates (\ref{rates}) exhibits particularly interesting dynamics starting from homogeneous initial conditions with a particle density $\rho >\rho_c$, which have first been described in \cite{godreche2003dynamics,godreche2005dynamics} for $\gamma =1$. The stages of the dynamics can be summarized as follows (cf.\ also Figure \ref{fig:fig1dynamic}):
\begin{enumerate}
\item Nucleation regime: the density decreases locally to $\rho_c$ and the resulting excess mass concentrates into cluster sites. Outside clusters on a so-called bulk sites, the system relaxes to its steady state distribution $\nu_{\phi_{c}}$.
\item Coarsening regime: the clusters exchange particles through the
bulk, leading to a decreasing number of cluster sites of increasing
size. This occurs on a slower timescale which is $L^{1+\gamma}$ for a complete graph or asymmetric system of size $L$ depending on the parameter $\gamma$, as derived section 4. In one-dimensional symmetric systems, the scale is $L^{2+\gamma}$.
\item Saturation Regime: on finite systems, eventually there
is only a single cluster site left which contains all excess particles and forms the condensate, which changes position on an even longer time scale. 
\end{enumerate}
Our main interest in this paper is the coarsening regime and we will explain in the following how to effectively describe the dynamics. We focus our presentation on the complete graph for simplicity, i.e. $q(x,y)=1/(L-1)$ for all $x\neq y$ with finite lattices $\Lambda$ of size $L$, and comment on the influence of other homogeneous geometries in the conclusion.

\section{Dynamics of the process}

\subsection{Empirical processes}

Consider a zero-range process with rates (\ref{rates}) on a finite lattice of size $L$ with $N$ particles, initially distributed uniformly at random. We denote the set of all possible configurations of this process by $\Omega_{L,N}=\{\boldsymbol{\eta}\in\Omega:\sum_{x\in\Lambda} \eta_x =N\}$. To describe the coarsening dynamics, we will study dynamics of two kinds of empirical processes:

\begin{eqnarray}
\mbox{Site empirical process}
&\quad F_{k}(\boldsymbol{\eta}(t)):=\frac{1}{L}\sum_{x\in\Lambda}\delta_{\eta_{x}(t),k}, \label{fk}
\end{eqnarray}

\begin{eqnarray}
\mbox{Size-biased empirical process}&\quad P_{k}(\boldsymbol{\eta}(t)):=\frac{1}{N}\sum_{x\in\Lambda}{k\delta_{{\eta_{x}(t)},k}}\ .\label{pk}
\end{eqnarray}

The first process counts the fraction of lattice sites with occupation number $k$, while the second one counts the fraction of particles which are on sites with occupation number $k$. Note that the condensed phase has only a small, time-dependent weight in the first process which tends to the stationary value $1/L$ with only one condensate site. In the second process, however, the weight of the condensed phase is roughly constant in time and given by the mass fraction $(\rho -\rho_c)/\rho$ due to the different weighting, where $\rho=N/L$ is the particle density. This is, therefore, more useful to describe the dynamics in the condensed phase. Both processes are normalized, i.e. $\sum_{k=0}^{\infty}F_{k}(\boldsymbol{\eta})=\sum_{k=0}^{\infty}P_{k}(\boldsymbol{\eta})=1$ for all $\boldsymbol{\eta} \in \Omega_{L,N}$. Also, they can be interpreted as distributions of single site occupations. The second process is a size-biased version of the first one, and both are related via
\begin{equation}\label{rel}
k F_{k}(\boldsymbol{\eta})=\rho P_{k}(\boldsymbol{\eta})\quad\mbox{for all }\eta\in\Omega_{L,N}\quad\mbox{and } k\geq 1.
\end{equation}

\begin{figure}[t]
\centering
\subfloat{\protect\includegraphics[scale=0.35]{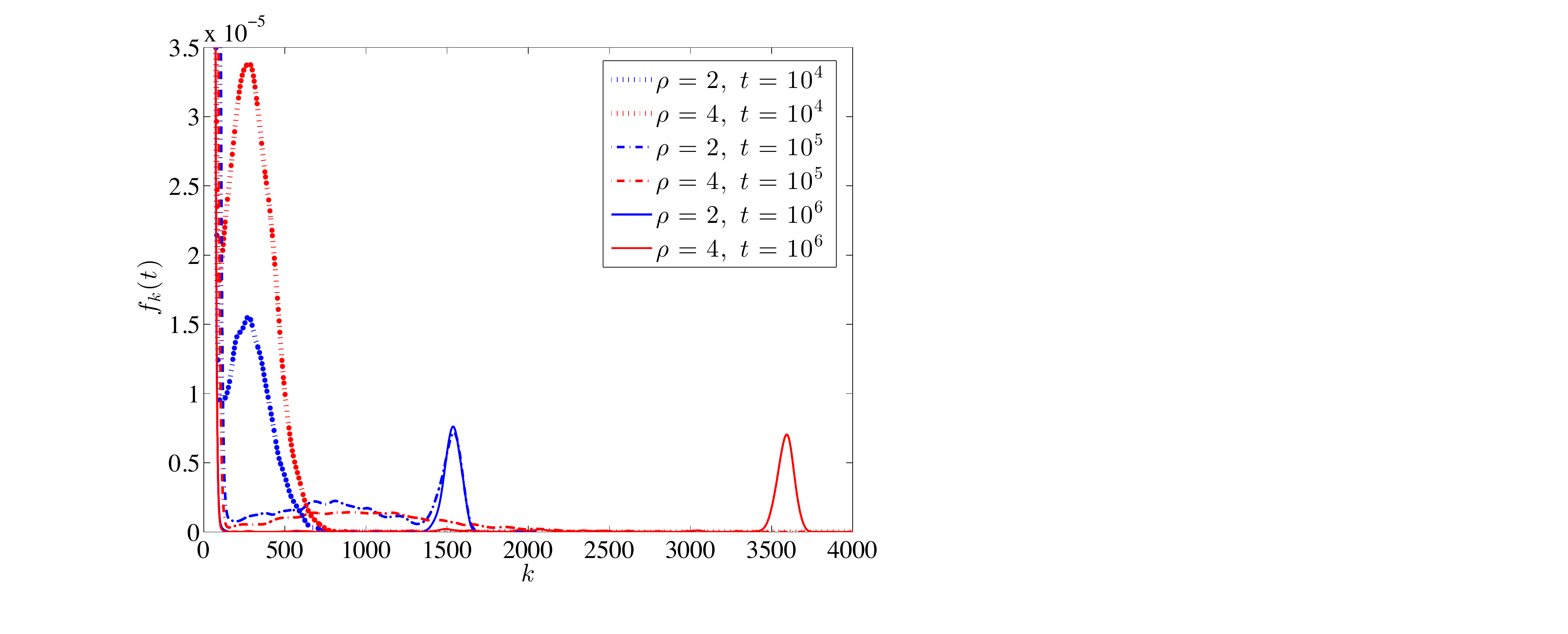}
}\subfloat{\protect\includegraphics[scale=0.36]{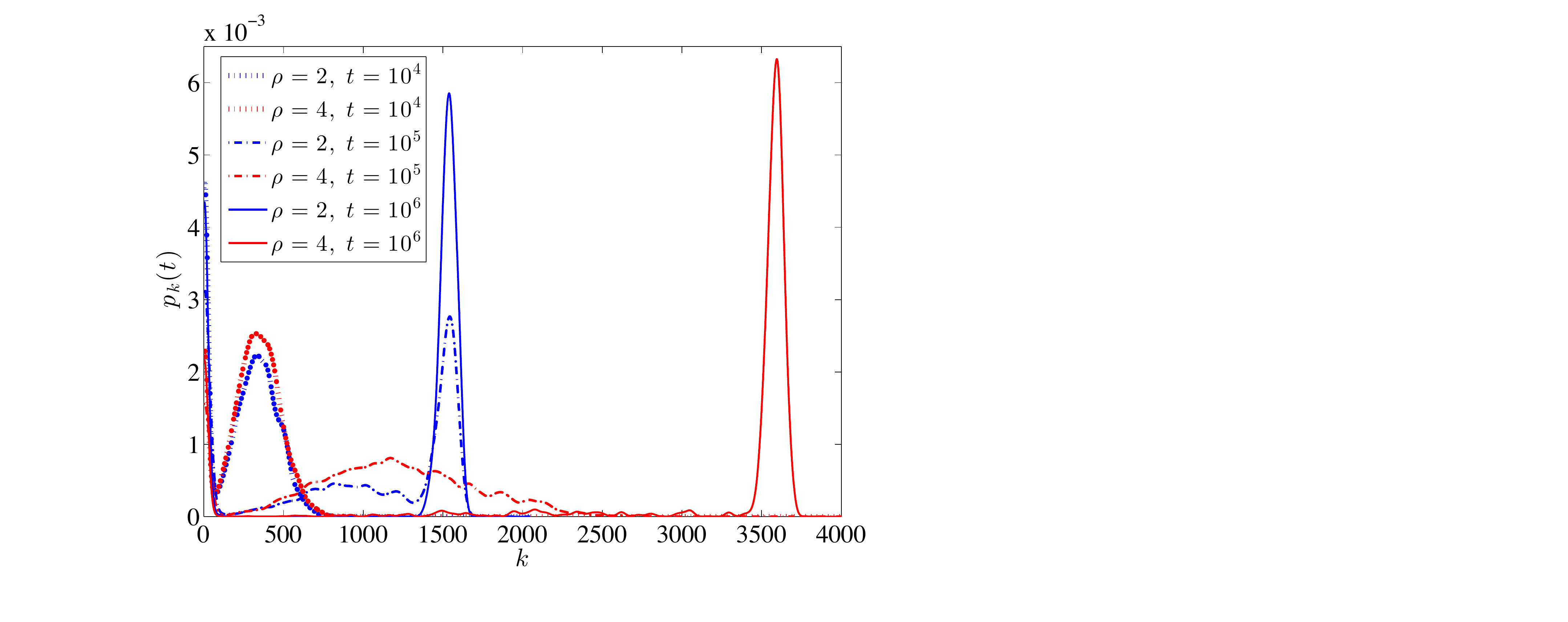}
}
\protect\caption{\label{fig:fkpk}Time evolution of the mass distribution as described by the empirical processes (\ref{fk}) and (\ref{pk}) for a system with $L=1024,\;\gamma =1,\;b=4$ and two different densities $\rho=2$ and $4$, which are larger than $\rho_c =0.5$. Expectations $f_k(t)$ (left) and $p_k(t)$ (right) are defined in (\ref{fke}) and (\ref{pke}). Outputs are at three different times, the first two during the coarsening regime, the last one when stationarity is reached. Note that the different scales of the $y$-axes indicate the larger weight of the conensed phase under the $P_k$ process. The bulk part of the distribution concentrates on the far left and is not resolved in this plot. Data are averaged over $500$ realisations.}
\end{figure}

In Figure \ref{fig:fkpk}, we illustrate the behaviour of these processes when averaged over $500$ realizations of the zero-range dynamics. We see that the coarsening process transports mass from small occupation numbers to form a stationary bump around occupation numbers $(\rho -\rho_c )L$, two such bumps are shown for densities $\rho =2$ and $4$. The distribution within the condensed phase at any given time is characterized by an intermediate bump in the distribution which is broadening over time with its maximum moving to the right. Building on the analysis in \cite{godreche2003dynamics}, we will analyze the scaling behaviour of this bump. On a finite lattice, the scaling behaviour only occurs in a finite time window before the system reaches stationarity. However, ideally, we are interested in an infinite lattice where the condensed phase distribution keeps broadening and moving to larger occupation numbers forever. Both processes in Figure \ref{fig:fkpk} are shown on very different scales since the condensed phase has much larger weight under $P_k$ than under $F_k$, as explained above. The bulk part of the distribution concentrates on the left and is not shown in full since we focus our attention on the condensed phase.

\subsection{Analysis of $F_{k}(\eta)$}

While the size-biased process $P_k$ is more suitable to study the coarsening dynamics, we start our analysis with the process $F_k$ describing the single site dynamics of the process since this is simpler to analyze. This has already been studied in \cite{godreche2003dynamics} for $\gamma =1$, so we keep the presentation short and present numerical results only for the size-biased process later in Section 4. 
Applying the generator (\ref{eq:GenZRP}) for the complete graph case with $q(x,y)=\frac{1}{L-1}$ to the test function $F_k$, we get

\begin{eqnarray*}
(\mathcal{L}F_k)(\boldsymbol{\eta}) 
&= \sum_{x, y\neq x} \frac{1}{L-1}g(\eta_x)[ F_k(\boldsymbol{\eta}^{x {\rightarrow} y}) - F_k(\boldsymbol{\eta})]\nonumber\\
&=-g(k)F_{k}(\boldsymbol{\eta})-\frac{1}{L-1}\sum_{x\in\Lambda\atop y\neq x}g(\eta_x)\frac{\delta_{k,\eta_{y}}}{L}\\
&\quad+\frac{1}{L-1}\sum_{x\in\Lambda\atop y\neq x} g(\eta_x)\frac{\delta_{k-1,\eta_{y}}}{L}+g(k+1)F_{k+1}(\boldsymbol{\eta})\nonumber\\
&=-(g(k)+\langle g\rangle_\eta )F_{k}(\boldsymbol{\eta})+\langle g\rangle_\eta F_{k-1}(\boldsymbol{\eta})+g(k+1)F_{k+1}(\boldsymbol{\eta})\\
&\quad+\frac{1}{L-1}\big( g(k)-\langle g\rangle_\eta \big)\big( F_{k}(\boldsymbol{\eta}) -F_{k-1}(\boldsymbol{\eta})\big)\ .\label{fkl}
\end{eqnarray*}

In the last line we have used the shorthand $\langle g\rangle_\eta :=\frac{1}{L}\sum_x g(\eta_x )$ for the spatial average of the jump rate. Since we are interested in the limit of large system size $L$, we will omit the correction term of order $1/(L-1)$ in the following. Now we use the evolution equation for expectations
\begin{equation}
\frac{d}{dt}\mathbb{E} \big[ F_k (\boldsymbol{\eta}(t) )\big] =\mathbb{E} \big[(\mathcal{L} F_k )(\boldsymbol{\eta}(t) )\big],
\label{evo}
\end{equation}
which is the analogue of the master equation for our particular observable $F_k$.

We use the notation
\begin{equation}
f_{k}(t)=\mathbb{E} \big [F_{k}(\boldsymbol{\eta} (t))\big],
\label{fke}
\end{equation}
and the shorthand
\begin{equation}
\langle g\rangle =\mathbb{E} \big [\langle g\rangle_{\eta (t)}\big] =\sum_{k=1}^\infty g(k) f_k (t),
\label{gmean}
\end{equation}
for the (time-dependent) expectation of the spatially averaged jump rate. Under the assumption that expections of products of the form $\langle g\rangle_\eta F_{k}(\boldsymbol{\eta})$ factorize, and that we can drop the correction terms of order $1/(L-1)$ as mentioned above, we end up with a closed equation
\begin{equation}
\frac{df_{k}(t)}{dt}=g(k+1)f_{k+1}(t)+\langle g\rangle f_{k-1}(t)-(g(k)+\langle g\rangle)f_{k}(t),\label{eq:df/dt}
\end{equation}
valid for all $k\geq 0$ with the convention $f_{-1}(t)\equiv 0$ for all $t\geq 0$.

We see that $\frac{d}{dt}\sum_{k}f_{k}(t)=0$, so that this can be interpreted as the master equation of a birth death chain with state space $\mathbb{N}_0$, birth rate $\langle g\rangle$ and state dependent death rate $g(k)$. In the following we denote the corresponding process by $(Y_t :t\geq 0)$, so that $f_k (t) =P[Y_t =k]$. Note that the birth rate is itself an expectation with respect to $f_k (t)$ (\ref{gmean}), so that this is in fact a nonlinear master equation. This chain describes the single-site dynamics of the zero-range process. With our choice of $\langle g\rangle$ in (\ref{gmean}) we see that
\begin{eqnarray}
\frac{d}{dt}\sum_k kf_k(t) 
&= \sum_k g(k+1)kf_{k+1}(t)+\langle g \rangle\sum_k kf_{k-1}(t) \nonumber \\
&\quad-\sum_k g(k)kf_k(t)-\langle g \rangle\sum_k kf_k \nonumber\\  
&= -\sum_k g(k)f_k(t)+{\langle g \rangle} =0\ ,
\label{masscon}
\end{eqnarray}
which is consistent with mass conservation and $\sum_{k}kf_{k}(t)=\rho$ for all $t\geq 0$. 
Due to this additional conserved quantity, the nonlinear chain has a whole family of stationary distributions, and indeed it is easy to see that the marginals $\nu_\phi (k):=\nu_\phi [\eta_x =k]$ (\ref{eq:marg}) of the zero-range process are stationary for all $\phi \in [0,1]$. For every initial distribution with density $\rho\leq\rho_c$, the chain will converge to the corresponding distribution with $R(\phi )=\rho$.

For supercritical densities $\rho>\rho_{c}$, we expect the chain to converge to $\nu_1$ in analogy to the behaviour of the zero-range process. The solution of (\ref{eq:df/dt}) will develop a bimodal structure over time, corresponding to the bulk and the condensed part of the distribution. This separation is characterized by a scaling parameter $\epsilon_t \to 0$ as $t\to\infty$, denoting the volume fraction of the condensed phase. Following the approach in \cite{godreche2003dynamics}, we therefore make the following ansatz for a separated state

\begin{equation}  
f_{k}(t)=\underbrace{f_k(t)\,\mathbb{I}_{[0,1/\sqrt{\epsilon_t}]} (k)}_{:= f^{\mathrm{bulk}}_k (t)}+\underbrace{f_k (t)\,\mathbb{I}_{(1/\sqrt{\epsilon_t},\infty)} (k)}_{:= f^{\mathrm{cond}}_k (t)}\ .
\label{ansatz}
\end{equation}

The split of the two contributions at occupation numbers $k=O(\epsilon_t^{-1/2})$ is arbitrary, both phases are clearly separated as can be seen later in Figure \ref{fig:BD sup}, and any other power in $(-1,0)$ would work equally well. The bulk part of the distribution converges i.e.
\begin{equation}
f^{\mathrm{bulk}}_k (t)\to \nu_1 (k)\quad\mbox{as }t\to\infty,
\label{bulkon}
\end{equation}
and in fact is very close to the stationary marginal $\nu_1$ even for finite times as is confirmed later in Figure \ref{fig:BD sub}. The probability fraction of this part is $(1-\epsilon_t) \to 1$ and the expectation converges to $\rho_c$ corresponding to the mass contained in the bulk.
The condensed part therefore contains the rest of the mass $\rho -\rho_c$ according to (\ref{masscon}), but only a vanishing probability of order $\epsilon_t$. Therefore, typical occupation numbers in this phase scale as $1/{\epsilon_t}$, which justifies the intermediate scale chosen in the formal definition (\ref{ansatz}). 
Furthermore, we assume that $f^{\mathrm{cond}}_{k}(t)$ takes the scaling form 
\begin{equation}
f^{\mathrm{cond}}_{k}(t)=\epsilon_{t}^{2}h(u),\label{eq:scaling_f}\quad\mbox{with }u=k\epsilon_{t}\mbox{ and }\epsilon_{t}=t^{-\frac{1}{\gamma+1}}\ ,
\end{equation}
which we adapted from \cite{godreche2003dynamics} to $\gamma\in(0,1]$. One of the prefactors $\epsilon_t$ corresponds to the vanishing probability of the condensed phase, the other one is a volume element to get a density $h(u)$. 
In terms of the scaling function $h(u)$, the volume fraction of the condensed phase is given by
\begin{equation}
\sum_{k}f^{\mathrm{cond}}_{k}(t)=\sum_{k \geq 1/\sqrt{\epsilon_t}}\epsilon_{t}^{2}h(k\epsilon_{t})\approx\epsilon_{t}\int_0^\infty h(u)du=\epsilon_t,
\label{eq:sum_fcon}
\end{equation}
which fixes the normalization $\int_0^\infty h(u)du=1$.
Also, the mass in the condensed phase is
\begin{equation}
\sum_{k}kf^{\mathrm{cond}}_{k}(t)=\sum_{k \geq 1/\sqrt{\epsilon_t}}\epsilon_{t}^{2}kh(k\epsilon_{t})\approx\int_0^\infty uh(u)du=\rho-\rho_{c},
\label{eq:moment_fcon}
\end{equation}
which is consistent with previous assumptions and imposes a second constraint on $h(u)$. 
Since the bulk phase is close to the stationary marginal $\nu_1$ and $h(u)$ changes with time only on scale $\epsilon_t$, we have
\begin{equation}
\frac{d}{dt}f^{\mathrm{bulk}}_k (t) =O(\dot{\epsilon_t})\ll O(\epsilon_t)=\frac{d}{dt}f^{\mathrm{cond}}_k (t)\ ,
\label{eq:order}
\end{equation}
as we will see later that $\epsilon_t$ decays as a power law with $t$. 
Therefore, the condensed part fulfills the same master equation as (\ref{eq:df/dt})
\begin{equation}
\frac{d}{dt}f^{\mathrm{cond}}_{k}(t)
=g(k+1)f^{\mathrm{cond}}_{k+1}(t)+\langle g\rangle f^{\mathrm{cond}}_{k-1}(t)-(g(k)+\langle g\rangle)f^{\mathrm{cond}}_{k}(t).
\label{eq:master_fcon}
\end{equation}
To write this as a closed equation for $h(u)$, we need to find an expression for the time-dependent birth rate $\langle g\rangle= \sum_{k\geq 1} g(k)\, f_k (t)$ under the phase separated state. Assuming local stationarity and using a balance of currents in both phases, we derive in Appendix A that
\begin{equation}
\langle g\rangle = 
1+A\epsilon_{t}^{\gamma} \label{eq:ansatz_gmean} \ ,
\end{equation}
where for $\rho$ large enough we expect $A=b/(\rho -\rho_c)^\gamma$. 
Plugging this into (\ref{eq:scaling_f}) and (\ref{eq:master_fcon}), we get for the leading order terms
\begin{equation}
{\frac{\dot{\epsilon_{t}}}{\epsilon_{t}}}[uh'(u)+2h(u)]=\epsilon_{t}^{\gamma+1}\left(\epsilon_{t}^{1-\gamma}h''(u)+\frac{b}{u^{\gamma}}h'(u)-Ah'(u)-\frac{b\gamma}{u^{\gamma+1}}h(u)\right).
\end{equation}
Equating powers of $\epsilon_t$ on both sides leads to the choice $\epsilon_{t}=t^{-\frac{1}{\gamma+1}}$ and the equation
\begin{equation}
t^{-\frac{1-\gamma}{1+\gamma}}h''(u)+\left(\frac{u}{(\gamma+1)}+\frac{b}{u^{\gamma}}-A\right)h'(u)+\left(\frac{2}{(\gamma+1)}-\frac{b\gamma}{u^{\gamma+1}}\right)h(u)=0,\label{eq:gamma_f}
\end{equation}
for the scaling function $h$. For $\gamma=1$, we simply have
\begin{equation}
h''(u)+\left(\frac{1}{2}u-A+\frac{b}{u}\right)h'(u)+\left(1-\frac{b}{u^{2}}\right)h(u)=0,\label{eq:gamma1_f}
\end{equation}
as was derived in \cite{godreche2003dynamics} which is indeed time-independent. For $\gamma <1$ we have to keep a small time-dependent prefactor for $h''$ to regularize the solutions. This is necessary since we impose conditions on the integral and first moment on $h$ as given in (\ref{eq:sum_fcon}) and (\ref{eq:moment_fcon}). Neither equation (\ref{eq:gamma_f}) nor (\ref{eq:gamma1_f}) can be solved explicitly, but numerical solutions presented in the next section agree well with simulation data.

\subsection{Analysis of $P_{k}(\eta)$}
Analogously to (\ref{fk}), we can act with the generator (\ref{eq:GenZRP}) on the test function
$P_{k}$ to get 
\begin{eqnarray*}
\centering
(\mathcal{L}P_{k})(\boldsymbol{\eta})
&=-\frac{k}{N}\sum_{x\in\Lambda}g(\eta_{x})\delta_{k,\eta_{x}}-\frac{k}{N(L{-}1)}\sum_{x\in\Lambda\atop y\neq x}g(\eta_{x})\delta_{k,\eta_{y}}\\
&\quad+\frac{k}{N(L{-}1)}\sum_{x\in\Lambda\atop y\neq x}g(\eta_{x})\delta_{k-1,\eta_{y}}+\frac{k}{N}\sum_{x\in\Lambda}g(\eta_{x})\delta_{k+1,\eta_{x}}.
\end{eqnarray*}
The first and last term can again be written as $-g(k)P_{k}(\boldsymbol{\eta})$ and $\frac{k}{k+1}g(k+1)P_{k}(\boldsymbol{\eta})$, respectively. For the two middle terms, we use the same approximation as before. Dropping terms of order $1/(L-1)$ we get
\begin{eqnarray}
(\mathcal{L}P_{k})(\boldsymbol{\eta})
&=-g(k)P_{k}(\boldsymbol{\eta})-\langle g\rangle_\eta P_{k}(\boldsymbol{\eta})+\frac{k}{k-1}\langle g\rangle_\eta P_{k-1}(\boldsymbol{\eta})\nonumber \\
&\quad +\frac{k}{k+1}g(k+1)P_{k+1}(\boldsymbol{\eta}),\label{eq:L}
\end{eqnarray}
for all $k\geq 1$ with the convention $P_{0}(\boldsymbol{\eta})/0 =0$. Note that by definition $P_{0}(\boldsymbol{\eta})=0$ since this is the size-biased version of the configuration. 
Taking expectations and using the same notation as before, with 
\begin{equation}
p_{k}(t)=\mathbb{E} P_{k}(\boldsymbol{\eta}(t)),
\label{pke}
\end{equation}
we get for $k=1$
\begin{eqnarray}
\frac{d}{dt}p_1(t)
&=-g(1)p_{1}(t)-{\langle g\rangle}p_{1}(t)+\frac{1}{\rho}{\langle g\rangle}f_{0}(t)+\frac{1}{2}g(2)p_{2}(t)\nonumber \\
&=\frac{1}{2}g(2)p_{2}(t)-2\langle g\rangle p_{1}(t)+\sum_{k\geq2}\frac{1}{k}(g(k)-\langle g\rangle)p_{k}(t),
\end{eqnarray}
where we used (\ref{gmean}) and that $f_{0}(t)=1-\sum_{k=1}^{\infty}f_{k}(t)=1-\rho\sum_{k=1}^{\infty}\frac{p_{k}(t)}{k}$. For $k>1$, we have
\begin{eqnarray}
\frac{d}{dt}p_k(t)
&=-g(k)p_k(t)-{{\langle g \rangle}}p_k(t)+\frac{k}{k{-}1}{{\langle g \rangle}}p_{k{-}1}(t)+\frac{k}{k{+}1}g(k{+}1)p_{k+1}(t)\nonumber\\ 
&=\frac{k}{k{+}1}g(k{+}1)p_{k+1}(t)+\frac{k}{k{-}1}{{\langle g \rangle}}p_{k-1}(t)\nonumber\\
&\quad-\left(\frac{k{-}1}{k}g(k)+\frac{k{+}1}{k}{\langle g \rangle}\right)p_k(t)+\frac{1}{k}({\langle g \rangle}-g(k))p_{k}(t).\label{eq:diffp_k}  
\end{eqnarray}
Again, this is the master equation of a birth death chain on the state space $\mathbb{N}=\{ 1,2,\ldots\}$ with additional diagonal terms corresponding to long-range jumps from occupation numbers $k>1$ to $k=1$, and killing or cloning events which are not probability conserving. We have
\begin{eqnarray}
\mbox{birth\;rate}&\quad &\frac{k+1}{k}{\langle g\rangle},\;\;\mbox{for}\;k>0\ ,\nonumber\\
\mbox{death\;rate}&\quad &\frac{k-1}{k}g(k),\;\;\mbox{for}\;k>1\ ,\nonumber\\
\mbox{rate\;from}\;k\;\mbox{to\;}1&\quad &\frac{1}{k}(g(k)-\langle g\rangle)_+ ,\;\mbox{for}\;k>1\ ,\nonumber\\
\mbox{cloning\;rate}&\quad &\frac{1}{k}(\langle g\rangle-g(k))_+ ,\;\mbox{for}\;k>1\ ,\nonumber\\
\mbox{killing\;rate}&\quad &\sum_{k>1}\frac{1}{k}(\langle g\rangle-g(k))_+ ,\;\mbox{for}\;k=1\ ,
\end{eqnarray}
where we denote by $(\cdot)_+ =\max\{ 0,(\cdot)\}$ the positive part of the expression. 
Note that the total cloning rate of chains with $k>1$ exactly equals the killing rate of chains with $k=1$, so in total the probability is conserved
\[
\frac{d}{dt}\sum_{k\geq 1}p_{k}(t)=\sum_{k\geq 2}\frac{p_{k}(t)}{k}(\langle g\rangle-g(k))+\sum_{k\geq 2}\frac{p_{k}(t)}{k}(g(k)-\langle g\rangle) =0.
\]
Note also that the average jump rate is now given by 
\begin{eqnarray}\label{gmeanp}
\langle g\rangle =\rho\sum_{k\geq 1} \frac{g(k)}{k} p_k (t),
\end{eqnarray}
using (\ref{rel}). If $\langle g\rangle <1$, all cloning and killing rates vanish and the only new part of the dynamics are long range jumps from $k$ to $1$. This leads to subcritical dynamics, and it is easy to see that now the size-biased version 
\begin{eqnarray}\label{sbmgn}
\bar\nu_\phi (k):= \frac{k}{R(\phi )} \nu_\phi [\eta_x =k]
\end{eqnarray}
of the marginals of stationary measures of the zero-range process (\ref{eq:marg}) are stationary for the birth death chain with master equation (\ref{eq:diffp_k}) for all $\phi\in [0,1]$. Note also that there is no obvious second conservation law for the size-biased chains related to the density as was the case for $f_k(t)$. However, $\rho$ now explicitly enters the master equation of the process through the above expression for $\langle g\rangle$ in (\ref{gmeanp}) which selects the stationary distribution for different $\rho \in [0,\rho_c]$. Any long jumps can be interpreted as sites with $\eta_x=0$ receiving a particle in the original zero-range processes. 
For supercritical systems with $\langle g\rangle\geq 1$, chains with small occupation number $k$ perform jumps to $1$ since $g(k)>\langle g\rangle$, while chains that made it to large occupation numbers do not but have a positive rate for cloning. This mechanism generates bimodal distributions with a condensed and a bulk phase by $p_k^{\mathrm{cond}}$ and $p_k^{\mathrm{bulk}}$, analogously to (\ref{ansatz}). 
Using the same scaling ansatz as (\ref{ansatz}) where we replace the asymptotic bulk part with a size-biased version, we note that $\rho p_{k}(t)=kf_{k}(t)$ implies the same relation for the condensate part of the distribution, i.e. $\rho p^{\mathrm{cond}}_{k}(t)=k f^{\mathrm{cond}}_{k}(t)$. 
This leads to
\[
\sum_{k}p^{\mathrm{cond}}_{k}(t)=\frac{1}{\rho}\sum_{k}k f^{\mathrm{cond}}_{k}(t)=\frac{\rho-\rho_{c}}{\rho}
\]
for the mass fraction in the condensed phase which does not vanish and is constant in time. In particular, for the scaling of $p^{\mathrm{cond}}_{k}(t)$, we have
\begin{equation}
p^{\mathrm{cond}}_{k}(t)=\frac{1}{\rho}k f^{\mathrm{cond}}_{k}(t)=\frac{1}{\rho}uh(u)\epsilon_{t}.
\end{equation}
Therefore, it is sufficient to solve the equations (\ref{eq:gamma_f}) or (\ref{eq:gamma1_f}) to get theoretical predictions for the scaling behaviour of $p^{\mathrm{cond}}_{k}(t)$.

\section{Main Results for the coarsening dynamics}
\subsection{Implementation of non-linear birth death chains}
Simulation of the birth death chains related to $f_k$ and $p_k$ can only be done approximately using a large ensemble of parallel realizations, due to the non-linearity of the master equations. To determine the time evolution, it is necessary to compute the time-dependent expectation $\langle g\rangle$, which can be approximated by an ensemble average. We denote the birth death chain related to $f_k$ by $(Y_{t}:t\geq 0)$ and let $Y^i_t$ be different realizations with $i=1,\ldots ,m$ in an ensemble of size $m$. With master equation (\ref{eq:df/dt}), we then use the approximation
\begin{equation}
\langle g\rangle \approx {\langle g\rangle}_m=\frac{1}{m}\sum_{i=1}^m g(Y^i_t )\ .
\end{equation}
We show in Appendix B that the total density in the ensemble average $\frac{1}{m}\sum_{i=1}^m Y^i_t$ is a martingale, i.e. future expectations are equal to its present value. As opposed to zero-range processes where the total number of particles is strictly conserved, this is still a fluctuating quantity. Furthermore, the ensemble has an absorbing state at $Y^i =0$ for all $i=1,\ldots m$, since the approximated birth rate in this state is ${\langle g\rangle}_m =0$. This leads to the fact that by fluctuations all ensembles get absorbed in state $0$, and one can show that the average time to absorption scales like $m$ (see Appendix B). This implies that the coarsening process can only be truthfully represented by the ensemble in a finite time window, which is similar to the restriction in the zero-range process due to the approach to stationarity. This represents a strong limitation for numerical analysis and is illustrated in Figure \ref{fig:sigma} further below. A further disadvantage is that in both processes, the chains related to $f_k$ and the original zero-range process, the condensed phase only covers a vanishing fraction of the ensemble or lattice, which leads to poor statistics as is illustrated in Figure \ref{fig:uh(u)} below. 

We can overcome these problems by simulating an ensemble of size-biased birth death chains $(X^i_t : t \geq 0)$ with master equation (\ref{eq:diffp_k}). The ensemble average
\begin{equation}
{\langle g\rangle}_m = \rho\sum_{i=1}^m g(X^i_t )/X^i_t\ ,
\label{eq:g_average}
\end{equation}
has to be modified as explained in the previous section since $X_t^i$ now represents a size-biased single site process. In contrast to the $Y_t^i$ chains, the density $\rho$ explicitly enters the dynamics as a parameter. In addition, we use the ensemble to implement killing and cloning events as follows: since the killing rate of chains with occupation number $k=1$ is equal to the total cloning rate for chains with $k>1$, everytime we clone a chain, we kill a chain with $k=1$ to keep the ensemble size $m$ fixed. There may be instances where no $k=1$ chain exists in the ensemble at the time of cloning, but it turns out that for large enough ensemble sizes this happens very rarely (for our parameter values not more than $10$ times in simulations up to times of order $10^6$), and we can ignore such events. They could easily be taken into account by allowing the ensemble size to grow in time, but this does not make any difference to numerical results.

\subsection{Main results}
\begin{figure}[t]
\centering
\subfloat{\protect\includegraphics[scale=0.60]{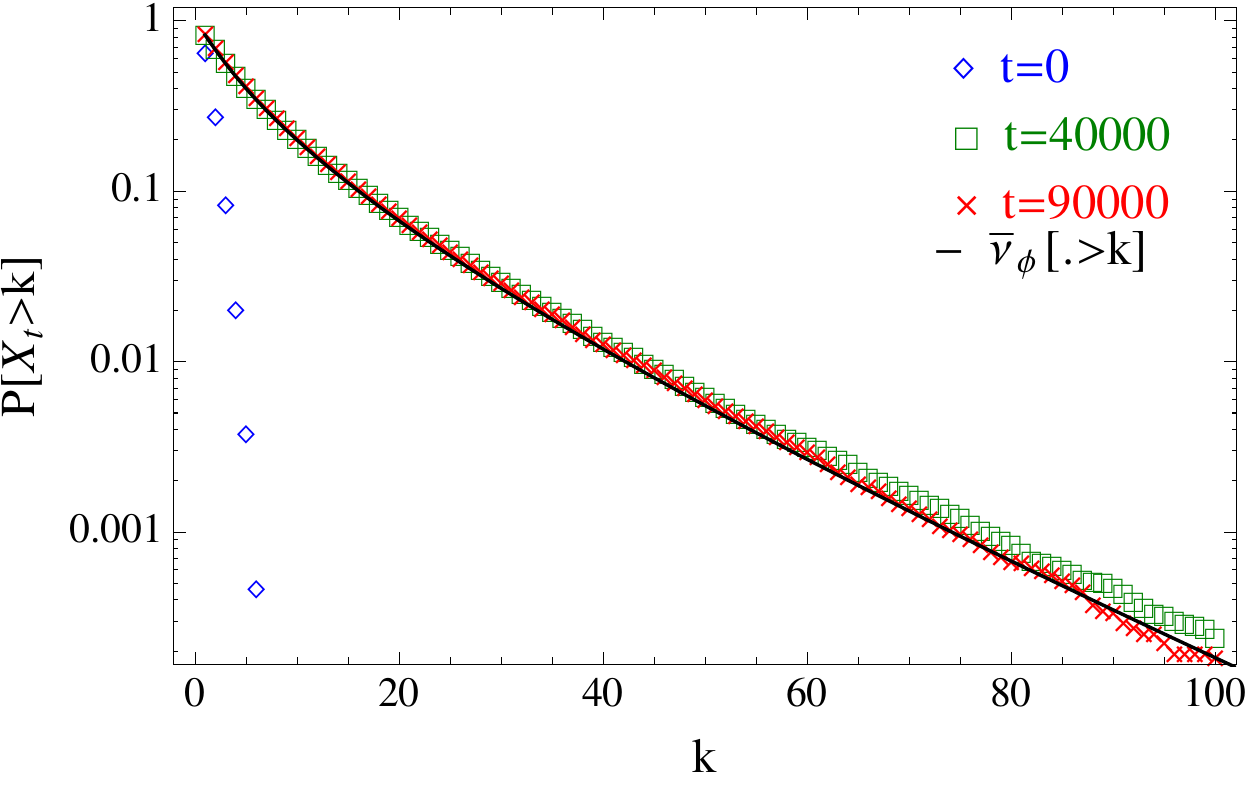}}
\subfloat{\protect\includegraphics[scale=0.60]{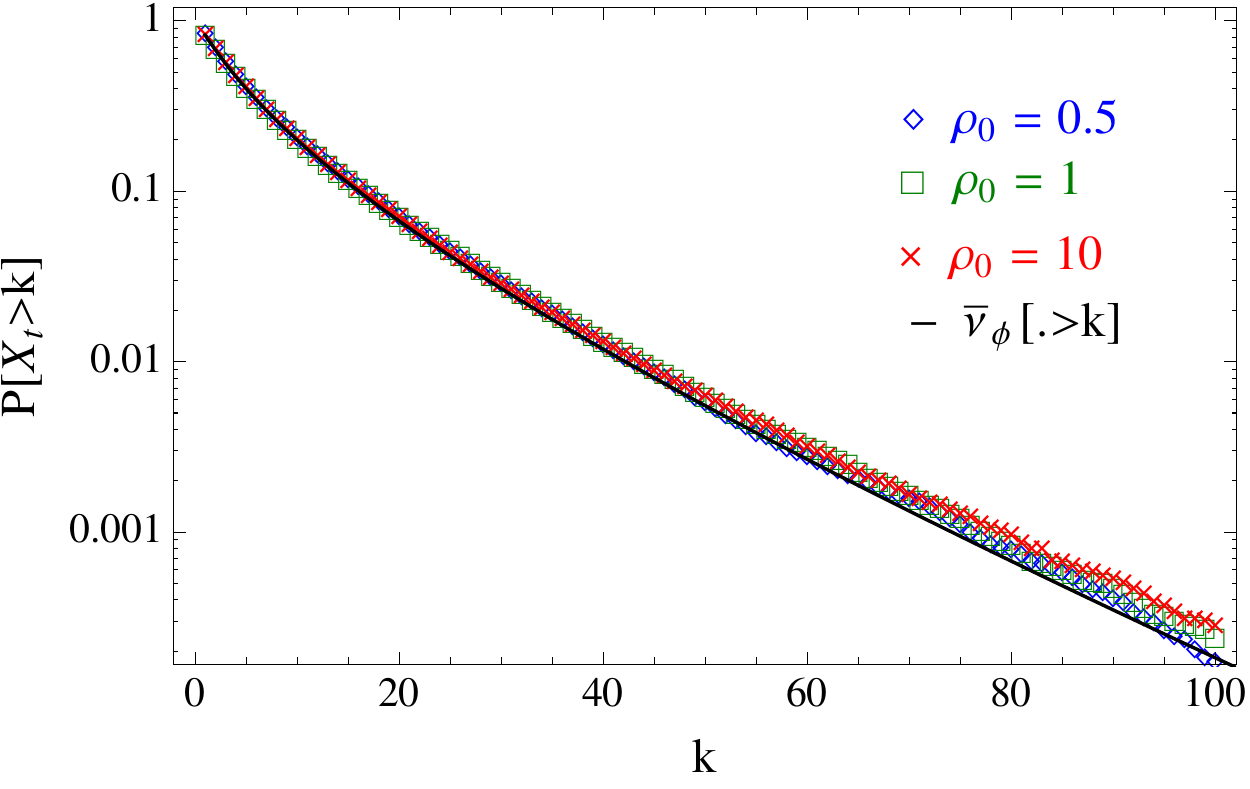}}
\protect\caption{\label{fig:BD sub}Convergence of the tail distribution for the size-biased chain $(X_t : t\geq 0)$ in the subcritical case $\rho<\rho_{c}$ to a size-biased version of the stationary distribution $\bar\nu_\phi$ (\ref{sbmgn}) with $\phi=0.95$ such that $R(\phi )=\rho$. The limiting distribution has an exponential tail, and the initial condition is Poisson with density $\rho_0$. Parameter values are $\gamma=1$ with $b=2.5$ and ensemble size $m=10^5$. The particle density is $\rho_0=\rho=1<\rho_c =1/(b-2)=2$ (left) while we also confirm that convergence only depends on the parameter $\rho=1$ entering the dynamics through (\ref{eq:g_average}), and is independent of the initial density $\rho_0 =0.5, 1$ and $10$ (right).}
\end{figure}
\begin{figure}[th]
\centering
\subfloat{\protect\includegraphics[scale=0.60]{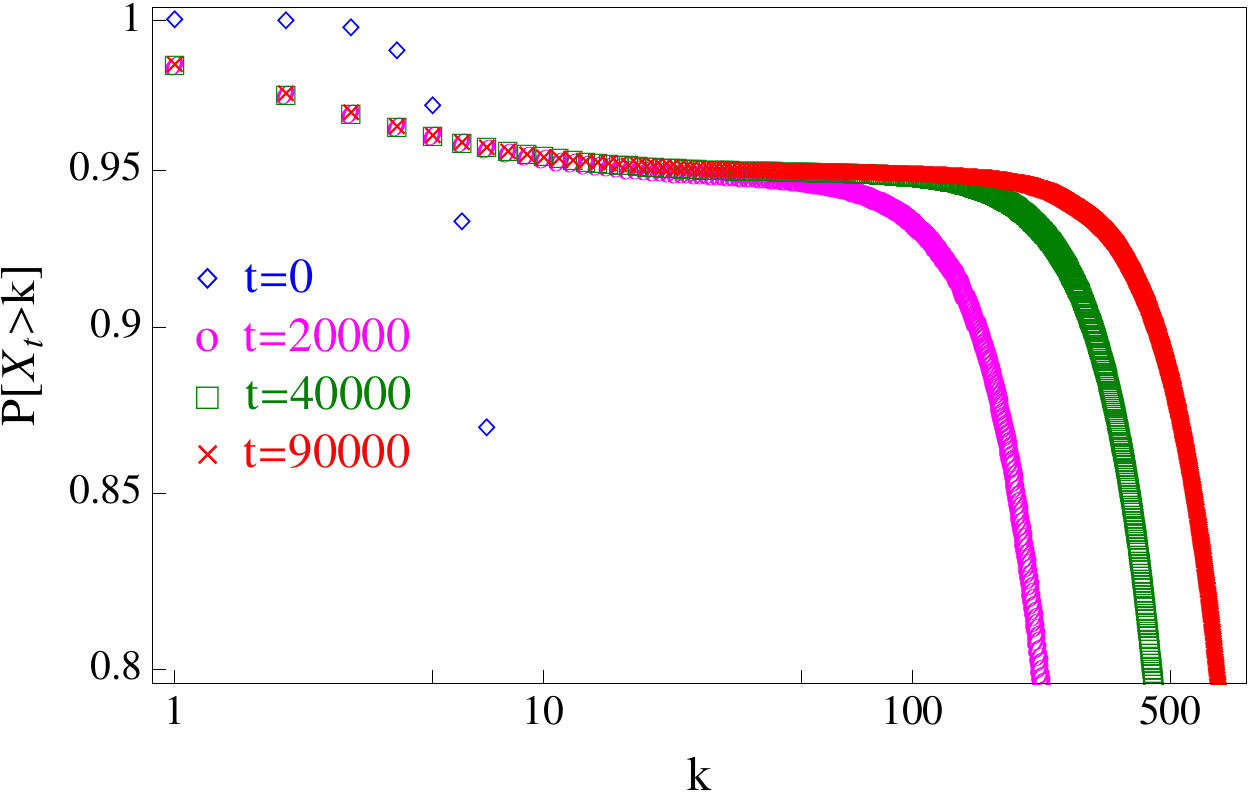}
}\subfloat{\protect\includegraphics[scale=0.60]{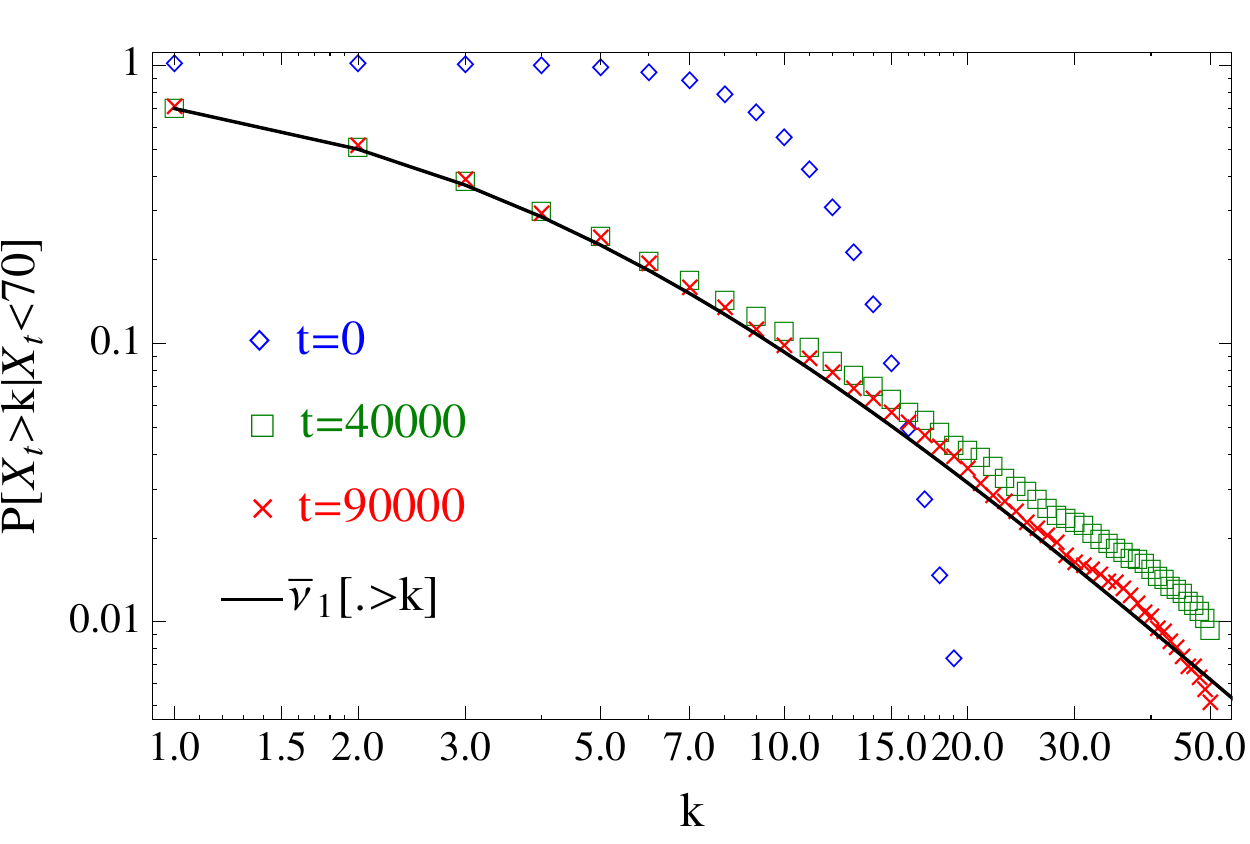}
}
\protect\caption{\label{fig:BD sup}Time evolution of the tail distribution of the size-biased chain $(X_t : t\geq 0)$ in the supercritical case $\rho >\rho_{c}$. Parameter values are $\gamma=1,\;b=4$ and $\rho=10>\rho_c =0.5$, the ensemble size is $m=10^5$. A plateau emerges at the value $1-\frac{\rho_{c}}{\rho}=0.95$ characterizing the phase separation as shown on the left. While the bulk part conditioned on small occupation numbers $\mathbb{P}[X_{t}>k\mid X_{t}<70]$ converges to the tail of a size-biased stationary marginal of the zero-range process $\bar\nu_1$ (\ref{sbmgn}) as shown in the right, the condensed part of the distribution keeps evolving to larger occupation numbers.}
\end{figure}
\begin{figure}[t]
\centering
\includegraphics[scale=0.4]{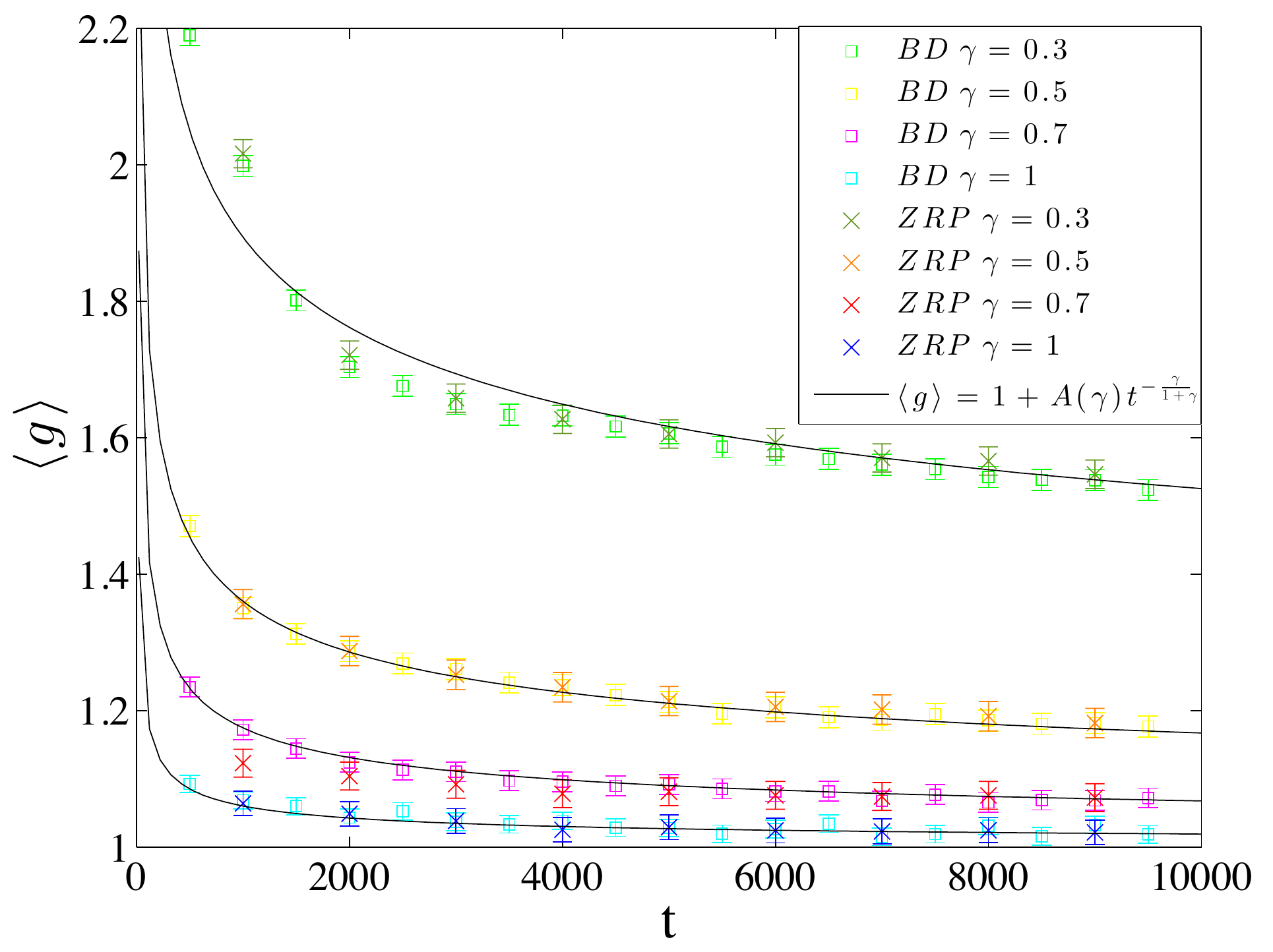}
\protect\caption{\label{fig:gmean}Agreement of the ensemble averages for the expected jump rate $\langle g\rangle$ under zero-range dynamics (ZRP) and the birth death chain $X_{t}$ (\ref{eq:g_average}) (BD), in comparison with the theoretical prediction (\ref{eq:ansatz_gmean}) with fitted parameters $A(\gamma )$. Parameter values are $b=4$ and $\rho=2$, system sizes are $L=m=1024$. The data for ZRP have further been averaged over $500$ realisations. Error bars denote standard error of mean and are comparable in both systems.}
\end{figure}

In the following we present our results comparing numerical solutions of the scaling equations with simulation results from the size-biased birth death chain with master equation (\ref{eq:diffp_k}), which we denote by $(X_{t}:t\geq0)$ where we have $X_{t}\in\mathbb{N}$. 
As a first test, we confirm that for subcritical densities $\rho<\rho_{c}$, $X_t$ eventually converges to a size-biased stationary distribution $\bar{\nu}$ as given in the previous section. This is illustrated in Figure \ref{fig:BD sub}, where the tail distribution of the process $X_t$ converges to a size-biased version $\bar\nu_\phi$ (\ref{sbmgn}) of the stationary marginal of the zero-range process. This is independent of the actual initial condition $X_0$, the asymptotic density $\rho$ is determined by the parameter $\rho$ in (\ref{gmeanp}) and (\ref{eq:g_average}) as shown in Figure \ref{fig:BD sub} (right). 
For supercritical $\rho>\rho_{c}$, the distribution of $X_t$ phase separates, where with probability $\frac{\rho_{c}}{\rho}$ it takes small values corresponding to the bulk sites of the zero-range process. This part of the distribution 
again converges to $\bar{\nu}_\phi$ with $\phi=1$ which has now a sub-exponential tail, as is illustrated in Figure \ref{fig:BD sup} (right). 
With probability $\frac{\rho-\rho_{c}}{\rho}$, the chain takes large values corresponding to the condensed phase. This is shown in Figure \ref{fig:BD sup} (left) where we plot the tail $P[X_{t}>k]$ and see a plateau emerging at $1-\frac{\rho_{c}}{\rho}$.
In Figure \ref{fig:gmean}, we compare the ensemble average for $\langle g\rangle$ under the ensemble of birth death chains with the spatial average in the zero-range process, and find good agreement. The fluctuations around the mean are of similar size as well since we choose $m$ and $L$ of similar size, even though much higher values of $m$ can be treated numerically without problems. Note that a single ensemble of $m=L$ chains gives the same quality of data as an average of $500$ realizations of the zero-range process of size $L$. The data also compare well with the theoretical prediction obtained in (\ref{eq:ansatz_gmean}) with a fitted constant $A$ depending on the parameter $\gamma$. 
\begin{figure}[th]
\centering
\subfloat{\protect\includegraphics[scale=0.4]{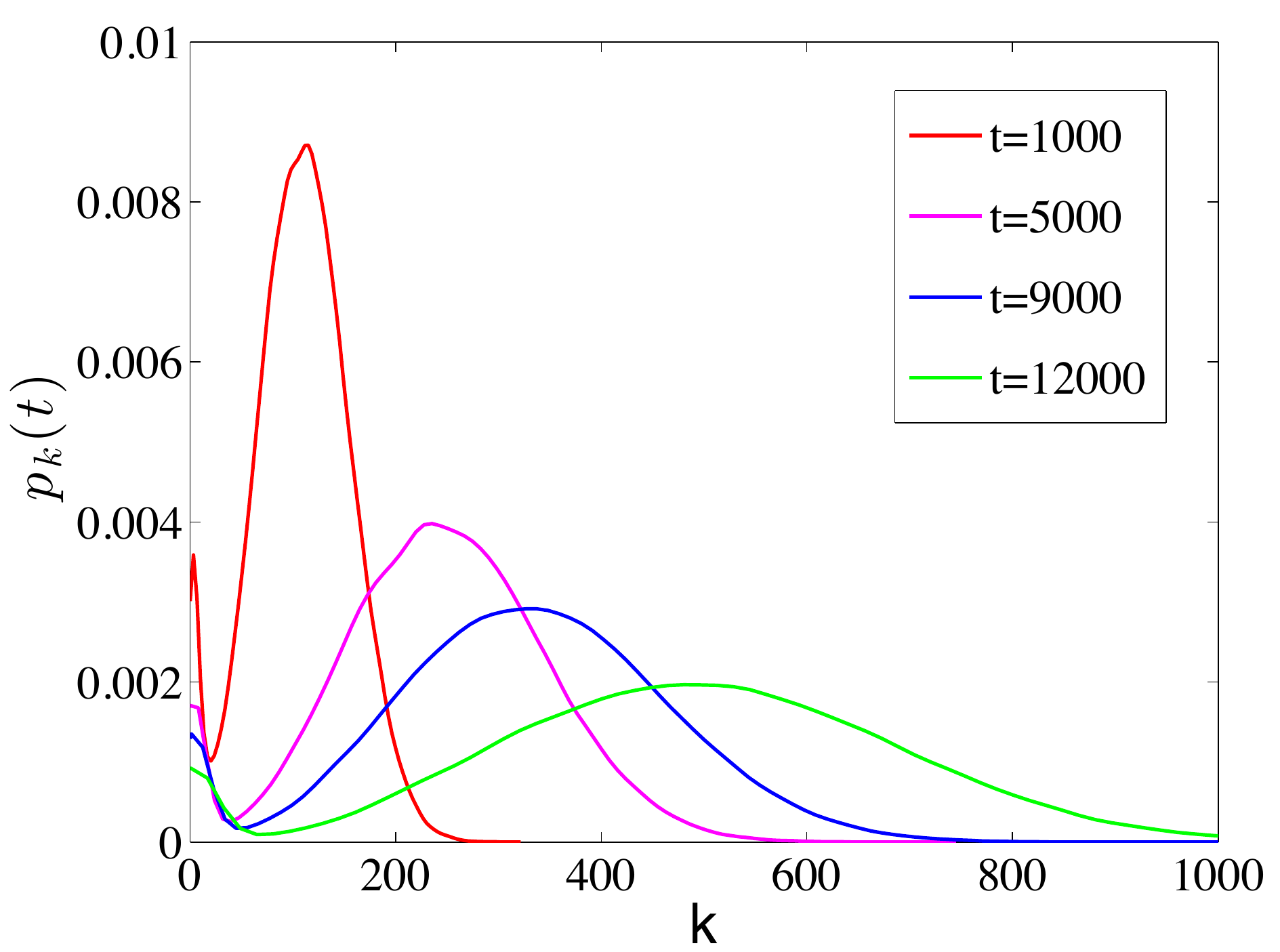}
}
\subfloat{\protect\includegraphics[scale=0.39]{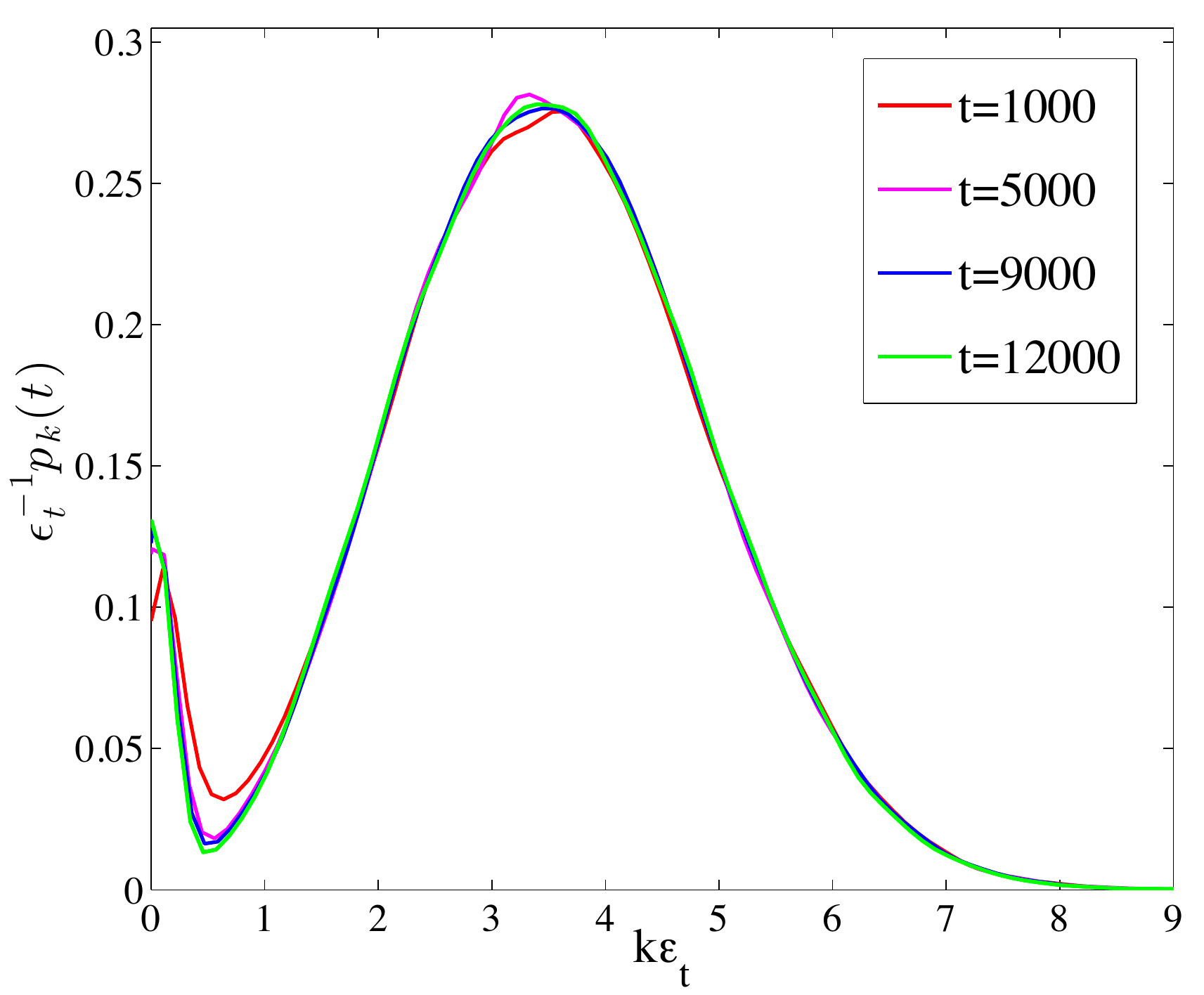}
}
\protect\caption{\label{fig:BD_p} Scaling behaviour of the condensed part of the distribution $p_{k}(t)$, with a data collapse when plotted against the rescaled variable $u=k\epsilon_t$ as shown in the right. Parameter values are $\gamma =1,\;b=4$ and $\rho=10$, with ensemble size $m=10^5$.}
\end{figure}
\begin{figure}[th]
\centering
\subfloat[$\gamma=0.5$]{\protect\includegraphics[scale=0.48]{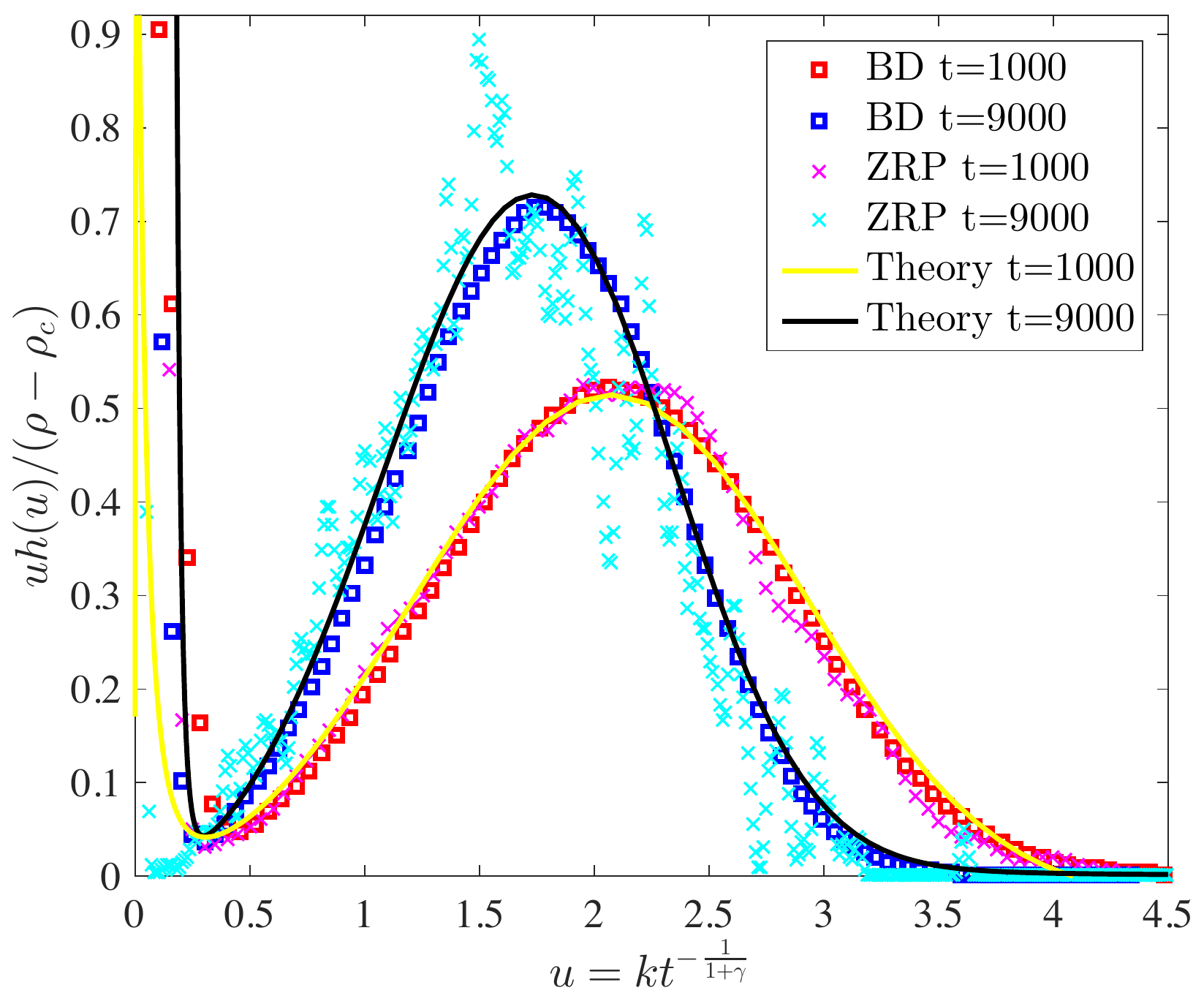}
}\subfloat[$\gamma=1$]{\protect\includegraphics[scale=0.4]{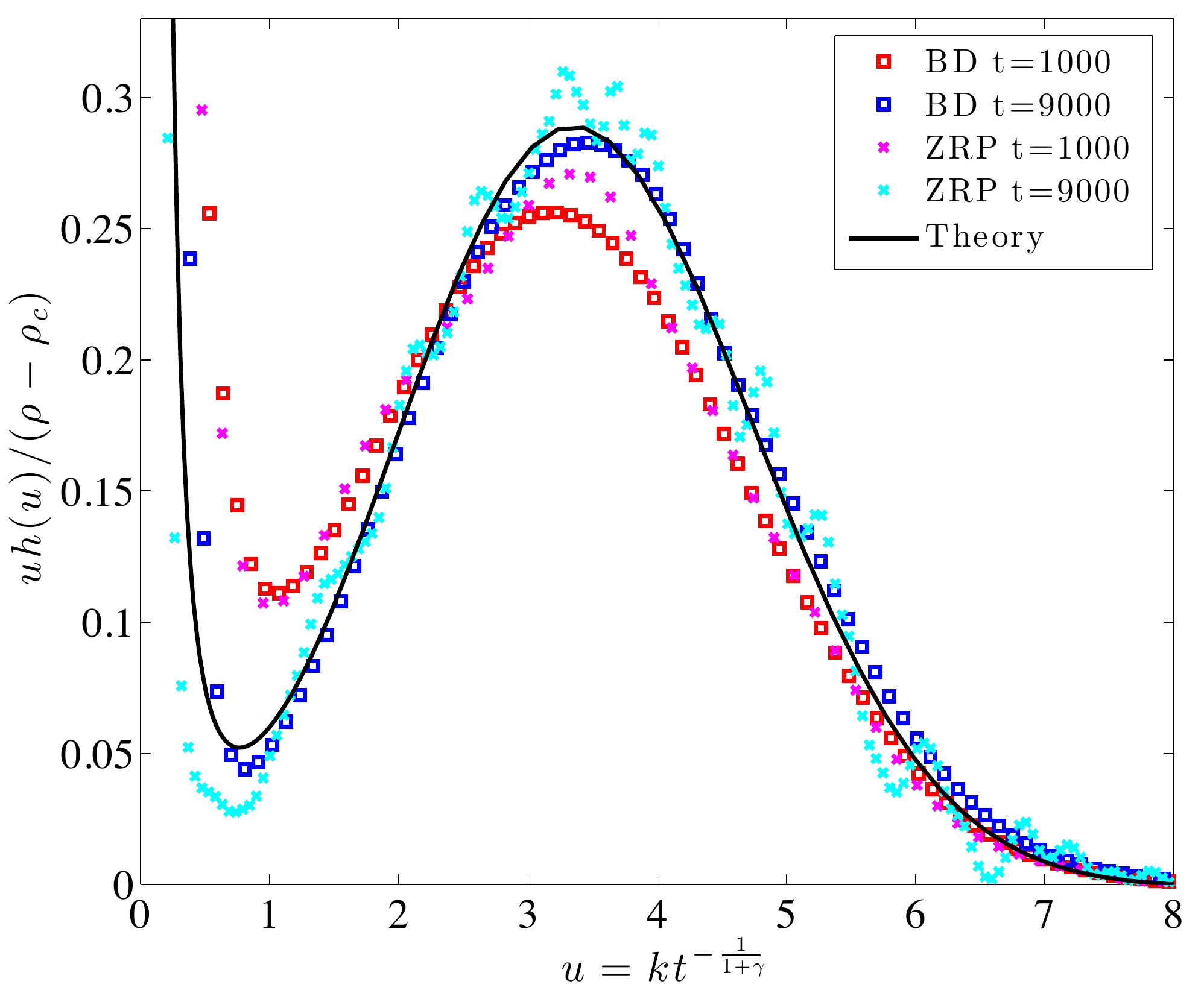}
}
\protect\caption{\label{fig:uh(u)}Normalized theoretical prediction $uh(u)$ as a solution of (\ref{eq:gamma_f}) and (\ref{eq:gamma1_f}) shows good agreement with asymptotic simulation data for the birth death chain $\epsilon_{t}^{-1}\rho p_{k}(\eta)$ (BD), plotted against the rescaled variable $u=k\epsilon_t$. Data from direct zero-range simulations (ZRP) coincide but are clearly of inferior quality in particular for large times since the volume of the condensed phase decreases. Parameter values are $b=4,\;\rho=2$ with $\gamma=0.5$ and $1$ and ensemble size $L=m=1024$. For ZRP, we further average over $500$ realizations while for BD we only use one. Note that for $t=\infty$ (\ref{eq:gamma1_f}) can be solved explicitly which is shown in (b).}
\end{figure}
By plotting the empirical distribution of $X_{t}$, we can see in Figure
\ref{fig:BD_p} (left) that the condensed part of the distribution has time independent mass $\frac{\rho-\rho_{c}}{\rho}$ and is moving to the right. Plotting against the rescaled occupation numbers $u=k\epsilon_t$ we see a data collapse confirming the predicted scaling in (\ref{eq:scaling_f}) as shown in Figure \ref{fig:BD_p} (right). 
The rescaled distributions of the condensed part also match well with the solution of (\ref{eq:gamma_f}) which is shown for $\gamma=0.5$ and $\gamma =1$ in Figure \ref{fig:uh(u)}. While for $\gamma=1$ the theoretical prediction is indeed independent of time, there is a time dependence for $\gamma<1$ as can be seen in the left plot. The behaviour as $t\to\infty$ for $\gamma<1$ is discussed in detail in \cite{godreche2016coarsening}, which appeared after the submission of this article. We also show data from a direct simulation of a zero-range process with the same numerical effort and bandwidth parameter for smoothing the density. The birth death chain obviously provides much better data for the condensed phase, and describes the distribution well also for relatively small time values.
\begin{figure}[t]
\centering
\subfloat{\protect\includegraphics[scale=0.35]{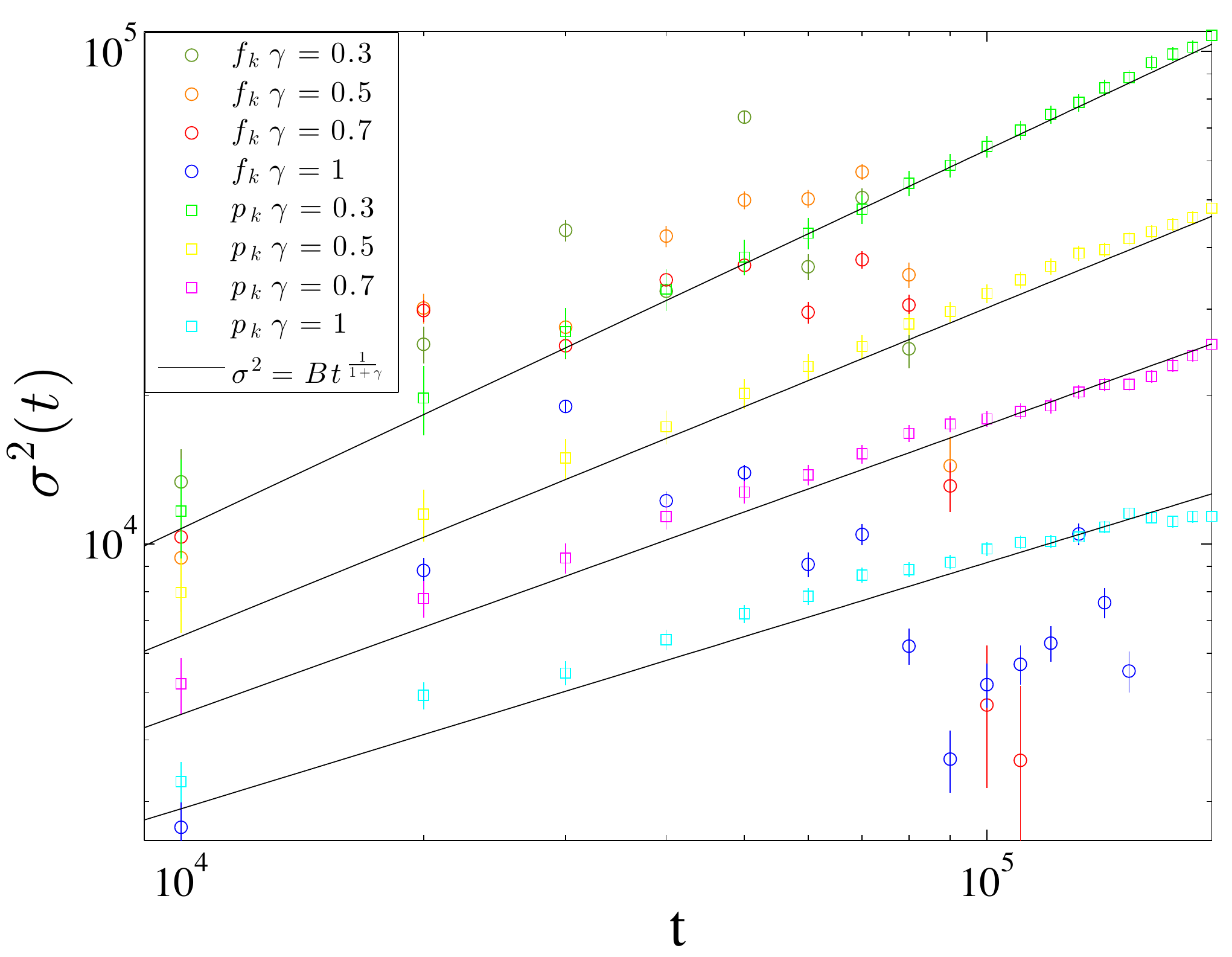}
}
\subfloat{\protect\includegraphics[scale=0.365]{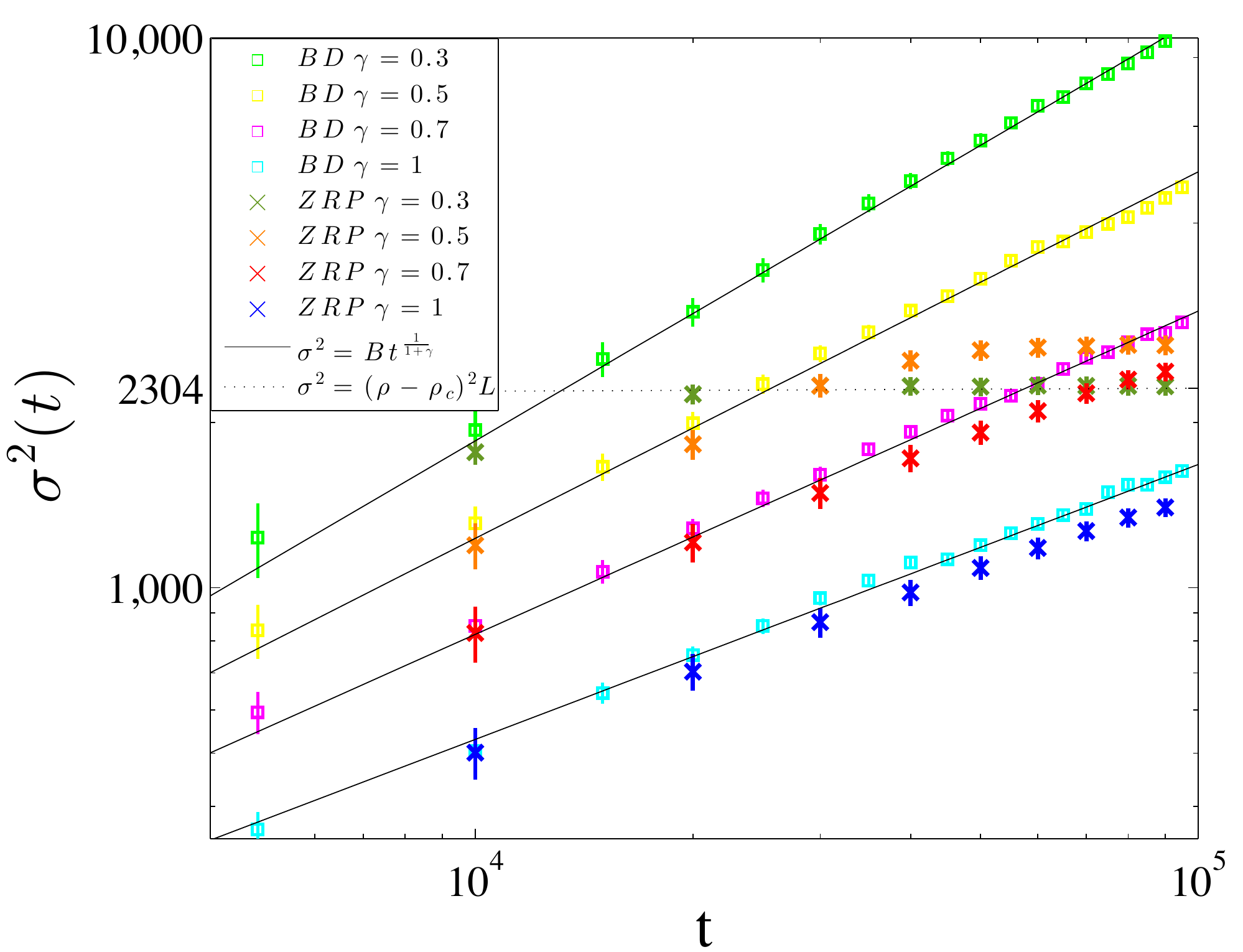}
}
\protect\caption{\label{fig:sigma}The second moment $\sigma^{2}(t)$ (\ref{defsigma}) increases asymptotically as a power law predicted in (\ref{eq:sigma}), as can be seen from simulation data of $X_{t}$ (shown by square boxes). We compare this to data from the birth death chains for $f_{k}(t)$ (left) with parameter values $b=4,\;m=1000$ and $\rho=10$, and to zero-range processes (right) with $b=4,\,\rho=2$ and system size $L=m=1024$. Both suffer from finite-size effects, the $f_k$ chains get eventually absorbed leading to large fluctuations at large times and the ZRP data saturate and converge to a stationary value around $(\rho-\rho_{c})^{2}L=2304$, which is visible for $\gamma=0.3$ and $0.5$ modulo finite size effects.
}
\end{figure}
Another interesting quantity is the second moment of the occupation numbers which can be computed as
\begin{equation}
\sigma^{2}(t)=\rho\mathbb{E}[p_{k}(t)]=\rho\sum_{k}kp_{k}(t)=\sum_{k}k^{2}f_{k}(t).
\label{defsigma}
\end{equation}
In the coarsening regime, the time evolution of $\sigma^{2}(t)$ is expected
to follow a scaling law analogous to results in \cite{godreche2003dynamics}. With the above expression, one can compute (see Appendix C)
\begin{equation}\label{eque}
\frac{d}{dt}\sigma^{2}(t)=\frac{d}{dt}\sum_{k\geq 1}k^2f_k(t)=2\rho(\langle g\rangle-1)+2\Big(\langle g\rangle-b\sum_{k\geq1}k^{1-\gamma}f_{k}(t) \Big),
\end{equation}
and for $\gamma=1$, this simplifies further to
\[
\frac{d}{dt}\sigma^{2}(t)=2\rho(\langle g\rangle-1)+2(\langle g\rangle-b(1-f_{0}(t)).
\]
Using that $\lim_{t\rightarrow\infty}f_{0}(t)=\nu_{1}[0]=\frac{b-1}{b}$, we get for large times
\begin{equation}
\frac{d}{dt}\sigma^{2}(t)=\Big(2\rho+2\Big)(\langle g\rangle-1)=\Big(2\rho+2\Big)A\epsilon_{t}^\gamma,
\end{equation}
which can be integrated to 
\begin{equation}
\sigma^{2}(t)=Bt^{\frac{1}{1+\gamma}},\label{eq:sigma}
\end{equation}
where $B=(2\rho+2)(1+\gamma)A$ is a constant related to $A$ and $\rho$. 
While exact for $\gamma=1$, for $\gamma <1$, explicit computations are not possible but numerical data strongly suggest that the second term in (\ref{eque}) does not affect the scaling of the second moment, and (\ref{eq:sigma}) remains valid when fitting constants $B$. 
This scaling law is plotted in Figure \ref{fig:sigma} which shows
good agreement to the simulation data of the size-biased chains. Data from the zero-range process agree in the coarsening regime, and data from the chains related to $f_k (t)$ show large fluctuations due to the existence of the absorbing state, so are numerically not very useful.

\section{Conclusion}
We have studied the coarsening dynamics towards condensation for zero-range processes with jump rates of the form (\ref{rates}) via the site and size-biased empirical processes, which are non-linear birth death chains with master equations (\ref{eq:df/dt}) and (\ref{eq:diffp_k}), respectively. We could close these equations with a standard mean field assumption and on a complete graph it should be possible to justify this rigorously which is left as future work. The coarsening time scale $\epsilon_{t}=t^{-1/1+\gamma}$ for $\gamma \in (0,1]$ is derived from a phase separated ansatz for the solution of the closed equation, analogously to results in \cite{godreche2003dynamics} for $\gamma=1$. The main novelty is the use of the size-biased birth death chain providing a strong tool to analyze the coarsening dynamics without finite size effects and significantly improved statistics. This approach is generic and can be adapted to other condensing particle systems such as the inclusion process, which is current work in progress.\\
Our results have been presented for the complete graph case. In other translation invariant geometries, such as one-dimensional lattices with periodic boundary conditions, the time scales can depend on local transport properties on the lattice, in particular on the symmetry of the dynamics.  
Asymmetric dynamics behave in general as the complete graph independently of the dimension. The scaling can be affected by the particular behaviour of first passage times and probabilities dominating the transport process of mass between cluster sites. This is slowing down coarsening for symmetric systems in one dimension with an expected scaling of $\sigma^2(t) \sim t^{\frac{1}{2+\gamma}}$ in analogy with results in \cite{godreche2003dynamics}, but in higher dimensions we expect the same scaling as for asymmetric systems and the complete graph, with logarithmic corrections in two dimensions. Mathematically this means that the mean field approximation to derive the master equations (\ref{eq:df/dt}) and (\ref{eq:diffp_k}) is not justified in certain symmetric systems and has to be adapted. In particular, the replacement of ${\langle g \rangle}_\eta=\frac{1}{2}(g(\eta_{x-1})+g(\eta_{x+1}))$ in (\ref{canocurrent}) in 1D symmetric systems has to be corrected since the presence of a cluster at site $x$ modifies the occupation number at sites $x-1$ and $x+1$, resulting in a decrease of the birth rates in the birth death chain. 

\section*{Acknowledgements}
The authors are grateful to Dario Span{\`o} for various helpful discussions. W. J. acknowledges funding from DPST, the Royal Thai Government scholarship.

\rhead{\emph{APPENDIX A}}
\section*{Appendix A. Current matching to derive $\langle g\rangle$ (\ref{eq:ansatz_gmean})}
\begin{figure}[t]
\centering
\subfloat{\protect\includegraphics[scale=0.6]{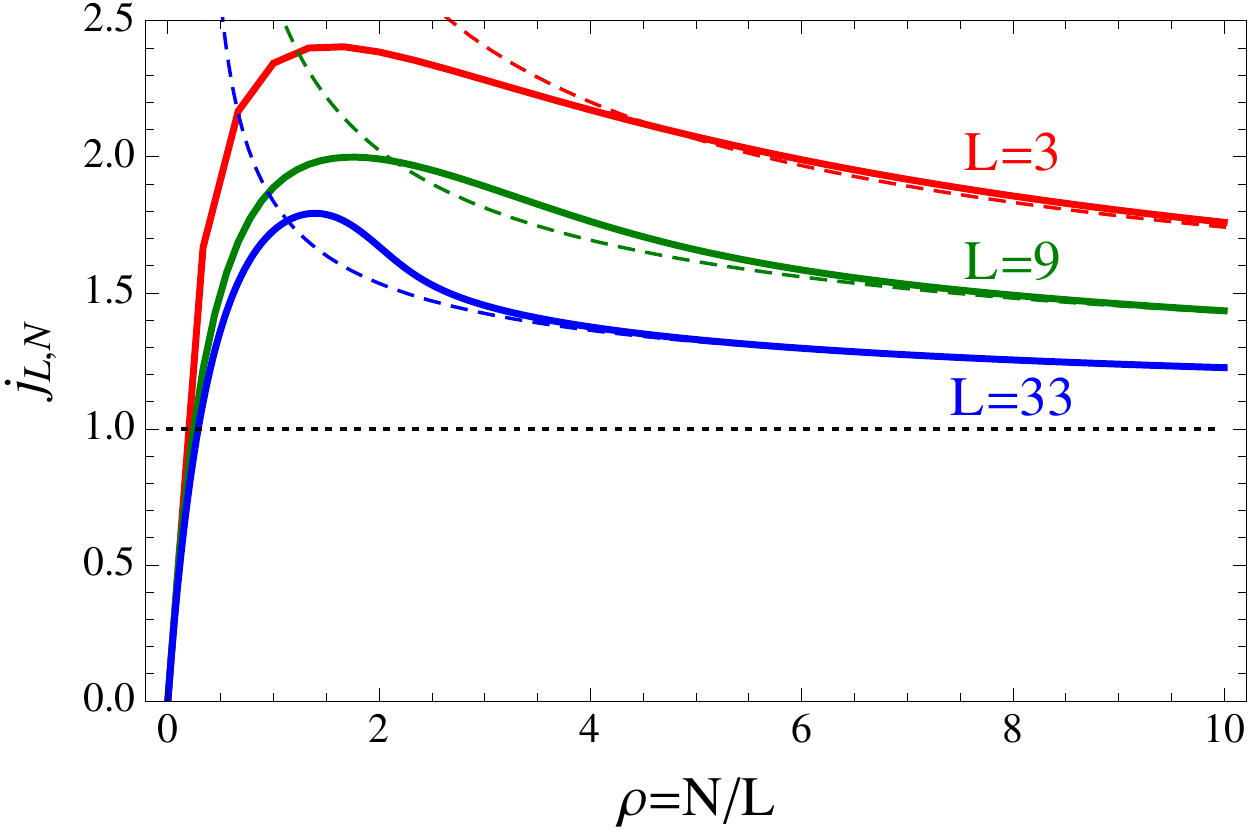}
}
\subfloat{\protect\includegraphics[scale=0.6]{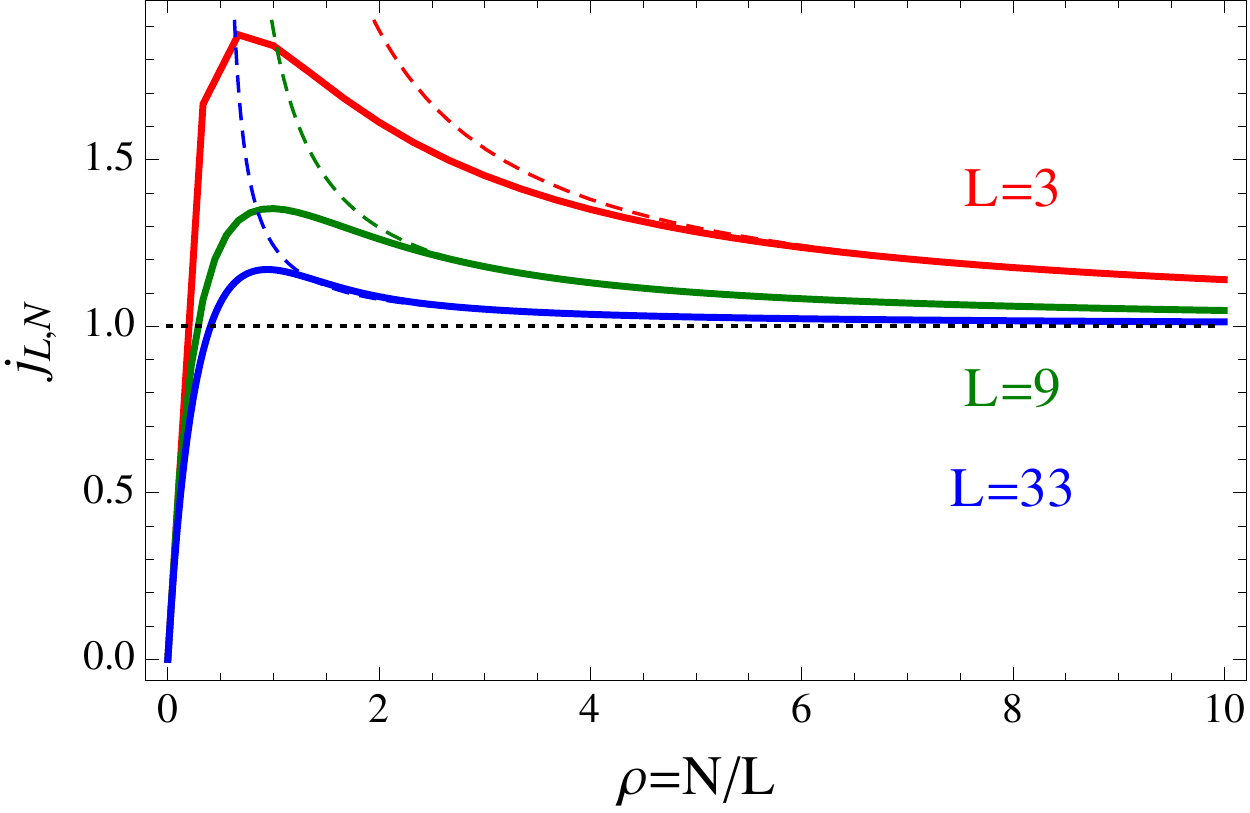}
}
\protect\caption{\label{fig:current}Approximation of the canonical current (\ref{canocurrent}) (bold line) to the jump rates (dashed line) as predicted in (\ref{currentjump}). Parameter values are $b=4$, $\gamma=0.5$ (left) and $\gamma=1$ (right).  
}
\end{figure}

On the subspace $\Omega_{L,N}=\{\eta \in \Omega_L: \sum\eta_x=N\}$, the zero-range process is ergodic, leading to unique stationary distributions given by 
\[
\pi_{L,N}[\boldsymbol{\eta}]:=\nu_\phi[\boldsymbol{\eta} \mid \sum_{x \in \Lambda} \eta_x =N].
\]
These canonical distributions are independent of $\phi$ and can be written as 
\[
\pi_{L,N}[\boldsymbol{\eta}]=\frac{\mathbb{I}_{X_{L,N}}(\boldsymbol{\eta})}{Z_{L,N}}\prod_{x \in \Lambda}w(\eta_x),
\]
where the canonical partition function is 
\[
Z_{L,N}=\sum_{\boldsymbol{\eta}\in \Omega_{L,N}}\prod_{x \in \Lambda}w(\eta_x).
\]
The expected jump rate off a site is proportional to the average stationary current in asymmetric systems or to the diffusivity in symmetric systems in the grand-canonical ensemble. In the grand-canonical ensemble, it is simply given by
\[
j_\phi := \mathbb{E}_{\nu_\phi}[g(\eta_x)]= \frac{1}{z(\phi)}\sum_{n=0}^{\infty}g(n)w(n)\phi^n=\phi,
\]
which can be calculated directly using the form of the stationary weights $w(n)$ given in (\ref{eq:marg}). In the canonical ensemble, the average current is simply given by a ratio of the partition functions
\begin{equation}
j_{L,N} := \mathbb{E}_{\pi_{L,N}}[g(\eta_x)]= Z_{L,N-1}/Z_{L,N}.
\label{canocurrent}
\end{equation}
This can be easily computed numerically using recursions of the form $Z_{L,N}=\sum_{n=0}^N w(n)Z_{L-1,n}$, and one finds that the supercritical current is well approximated by the jump rates as
\begin{equation}
j_{L,\rho L} \approx g((\rho -\rho_c )L) = 1+\frac{b}{(\rho -\rho_c )^\gamma L^\gamma} .
\label{currentjump}
\end{equation}
This is illustrated in Figure \ref{fig:current}, and this approximation works well even for very small system sizes. Now assume that $1/\epsilon_t$ is the typical separation between cluster sites, and the system is locally stationary in between clusters, so the main assumption is that the phase separated state in the limit of diverging system size is given by
\[
f_k (t)=\pi_{1/\epsilon_t ,\rho /\epsilon_t},
\]
i.e. the canonical, super-critical state with time-dependent system size corresponding to the typical distance between clusters. 
So we can replace $L$ by $1/\epsilon_t$ in (\ref{currentjump}) to get the prediction
\[
\langle g\rangle =1+\epsilon_t^\gamma \frac{b}{(\rho -\rho_c )^\gamma },
\]
which works best for large values of  $\rho$.

\rhead{\emph{APPENDIX B}}
\section*{Appendix B. Martingale property of the ensemble of birth death chains}
\begin{figure}[t]
\centering
\includegraphics[scale=0.45]{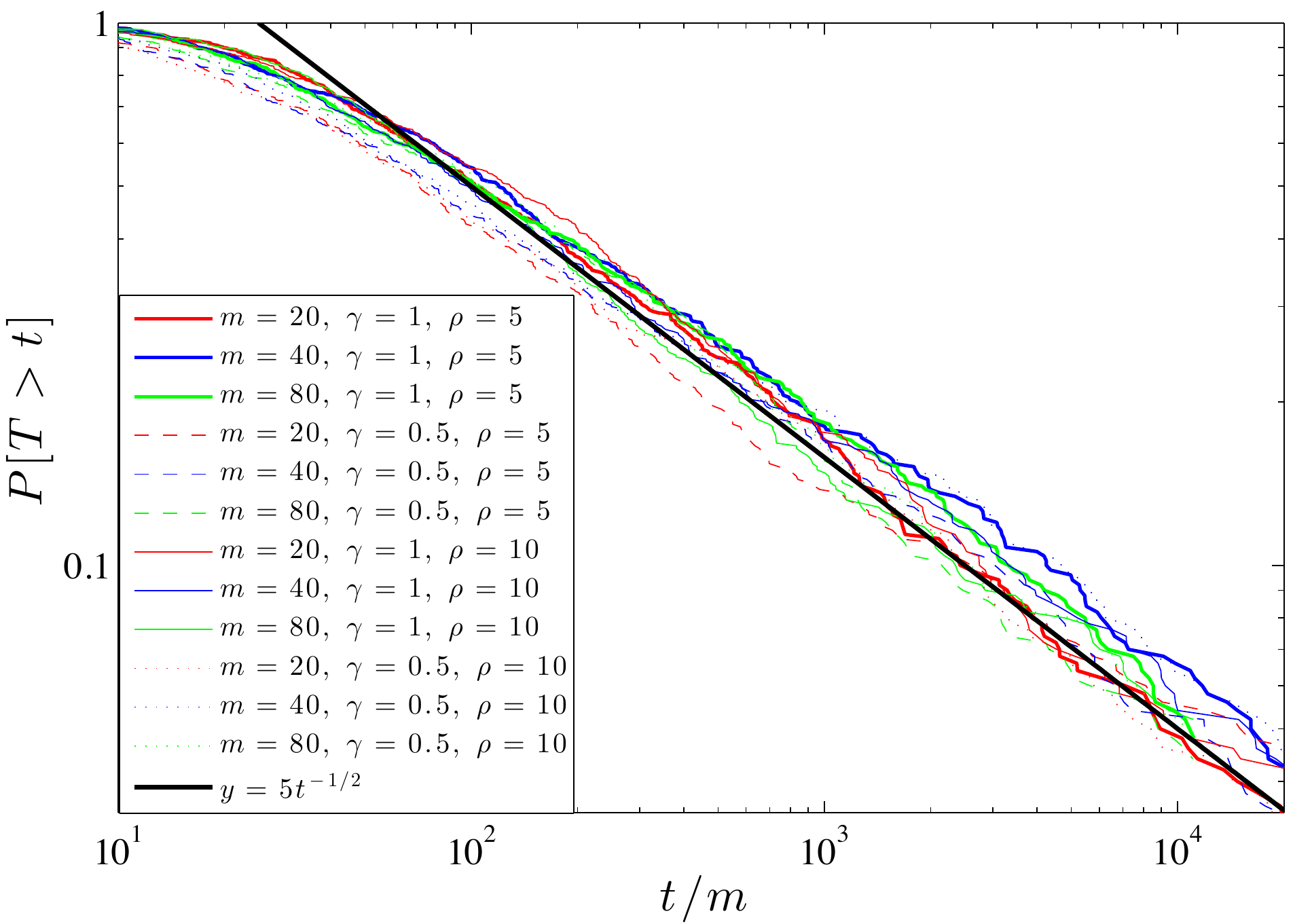}
\protect\caption{\label{fig:abstime} The tail distribution of the absorption time $T$ of $X_t$ for $b=4$ against rescaled time $t/m$ confirms the scaling of $T$ being linear in $m$. The actual distribution of $T$ has the same tail proportional to $t^{-1/2}$ as simple random walks or Brownian motion.}
\end{figure}
The dynamics of $m$ birth death chains $\boldsymbol{Y}=(Y^i:i=1,...,m)$ related to $f_k(t)$ is defined by the generator
\[
\mathcal{L}F(\boldsymbol{Y})=\sum_{i=1}^m g(Y^i)[F(\boldsymbol{Y}-\boldsymbol{e}^i)-F(\boldsymbol{Y})]+{\langle g \rangle}[F(\boldsymbol{Y}+\boldsymbol{e}^i)-F(\boldsymbol{Y})],
\]
where $\boldsymbol{e}^i\in\{0,1\}^m$ is the unit vector $({e}^i)_j=\delta_{ij}$.
Using the special test function $H(\boldsymbol{Y})=\sum_{i=1}^{m}Y^{i}$ for the total number of particles in the birth death chains, we get
\[
\mathcal{L}H(\boldsymbol{Y})=-\sum_{i=1}^{m}g(Y^{i})+\sum_{i=1}^{m}\frac{1}{m}\sum_{j=1}^{m}g(Y^{j})=0.
\]
This implies that $H(\boldsymbol{Y}_{t})=\sum_{i=1}^{m}Y_{t}^{i}$ is a (non-negative) martingale and in particular that $\mathbb{E}[H(\boldsymbol{Y}_t)]=\mathbb{E}[H(\boldsymbol{Y}_0)]$ for all $t \geq 0$.
Considering, $H(\boldsymbol{Y})^{2}=(\sum_{i=1}^{m}Y^{i})^{2}$ we get
\begin{eqnarray*}
\mathcal{L}(H^2)
&= \sum_{i=1}^m g(Y^i)[(H-1)^2-H^2]+{\langle {g} \rangle}[(H+1)^2-H^2]\\ 
&= \sum_{i=1}^m g(Y^i)[-2H+1]+\sum_{i=1}^m{\langle {g} \rangle}[2H+1]\\
&= -2m{\langle {g} \rangle}\sum_{i=1}^m Y^i +m{\langle {g} \rangle}+2m{\langle {g} \rangle}\sum_{i=1}^m Y^i +m{\langle {g} \rangle}\\
&= 2m{\langle {g} \rangle}.
\end{eqnarray*}
Therefore, the quadratic variation is $[H]_t=\int_0^t \mathcal{L}H^2(\boldsymbol{Y}_s)ds=2m\int_{0}^{t}\langle g\rangle_{s}ds$, which characterizes the fluctuations of the martingale $H(\boldsymbol{Y}_t)$ (see e.g. Chapter 2 in \cite{Ethier1986} for details). This is linear in $t$ to leading order as long as $\langle g\rangle>0$ is bounded. With $H(\boldsymbol{Y}_0)=\rho m$, the process can get absorbed in $\boldsymbol{0}$ when the variance reaches a level of order $(\rho m)^2$. Since $\langle{g}\rangle_s$ converges to $\phi(\rho)$, this implies $2mT\phi(\rho)=(\rho m)^{2}$ to characterize the absorption time $T$ for large $m$, which is then approximately $m\frac{\rho^{2}}{2\phi(\rho)} \approx m$.

\rhead{\emph{APPENDIX C}}
\section*{Appendix C. Derivation of the behaviour of $\sigma^2(t)$ (\ref{eque})}
With $\sigma^2(t)=\sum_k k^2f_k(t)$ as given in (\ref{defsigma}) and using the master equation (\ref{eq:df/dt}) we have
\begin{eqnarray*}
\frac{d}{dt}\sum_k{k^2f_k(t)}
&=\sum_k{g(k+1)k^2f_{k+1}(t)}+\sum_k{\langle g \rangle}k^2f_{k-1}(t)\\
&\quad-\sum_k(g(k)+{\langle g \rangle})k^2f_k(t)\\ 
&=\sum_{k\geq 1}{g(k)(k-1)^2f_k(t)}+\sum_k{\langle g \rangle}(k+1)^2f_k(t)\\
&\quad-\sum_k(g(k)+{\langle g \rangle})k^2f_k(t)\\
&=\sum_{k\geq 1}{g(k)(-2k+1)f_k(t)}+\sum_k{{\langle g \rangle}(2k+1)f_k(t)}\\
&=-2\sum_{k\geq1}{\left(1+\frac{b}{k^\gamma}\right)kf_k}+\sum_k{g(k)f_k(t)}\\
&\quad+2{\langle g \rangle}\sum_k{kf_k(t)}+{\langle g \rangle}\sum_k{f_k(t)}\\
&=-2\sum_{k\geq1} kf_k(t)-2b\sum_{k\geq1}{k^{1-\gamma}f_k(t)}+2{\langle g \rangle}\rho+2{\langle g \rangle}\\
&=-2\rho+2{\langle g \rangle}+2\rho{\langle g \rangle}-2b\sum_{k\geq1}{k^{1-\gamma}f_k(t)}.
\end{eqnarray*}
Therefore, we have
\begin{equation}
\frac{d}{dt}\sigma^{2}(t)=2\rho(\langle g\rangle-1)+{2}\Big(\langle g\rangle-b\sum_{k\geq1}k^{1-\gamma}f_{k}(t) \Big),
\end{equation}
as shown in (\ref{eque}).

\rhead{\emph{REFERENCES}}

\end{document}